\AtBeginDocument{%
  }

 \documentclass[acmsmall, screen]{acmart}

\DeclareMathDelimiter{(}{\mathopen} {operators}{"28}{largesymbols}{"00}
\DeclareMathDelimiter{)}{\mathclose}{operators}{"29}{largesymbols}{"01}

\usepackage{mathtools}  
\usepackage{paralist}
\usepackage{multirow}

\usepackage[table]{xcolor}
\usepackage{multicol}

\definecolor{lightblue}{RGB}{220,235,255}
\definecolor{lightred}{RGB}{255,220,220}

\usepackage{graphicx}

\usepackage[font=small]{caption}
\usepackage{subcaption}
\usepackage{bbm}

\usepackage{enumitem}
\usepackage{algorithm}
\usepackage{algpseudocode}
\usepackage{caption} 

\algrenewcommand\algorithmicrequire{\textbf{Input:}}
\algrenewcommand\algorithmicensure{\textbf{Output:}}

\definecolor{myred}{RGB}{220,43,25}
\definecolor{mygreen}{RGB}{0,139,0}
\definecolor{myblue}{RGB}{0,0,139}

\usepackage{amsthm}

\theoremstyle{definition}

\usepackage{wrapfig}
\begin{document}


\title{A Vertical Look at UAV Connectivity in the Wild: Cellular vs. Starlink, 3D Characterization, and Performance Prediction}

\author{Sravan Reddy Chintareddy}
\email{sravan.ch@ku.edu}
\affiliation{%
  \institution{University of Kansas}
  \streetaddress{1450 Jayhawk Blvd}
  \city{Lawrence}
  \state{Kansas}
  \country{USA}
  \postcode{66045}
}

\author{Sherwan Jalal Abdullah}
\email{sherwan.abdullah@ku.edu}
\affiliation{%
  \institution{University of Kansas}
  \streetaddress{1450 Jayhawk Blvd}
  \city{Lawrence}
  \state{Kansas}
  \country{USA}
  \postcode{66045}
}

\author{Justin D. Clough}
\email{justinclough@ku.edu}
\affiliation{%
  \institution{University of Kansas}
  \streetaddress{1450 Jayhawk Blvd}
  \city{Lawrence}
  \state{Kansas}
  \country{USA}
  \postcode{66045}
}

\author{Victor S. Frost}
\email{sravan.ch@ku.edu}
\affiliation{%
  \institution{University of Kansas}
  \streetaddress{1450 Jayhawk Blvd}
  \city{Lawrence}
  \state{Kansas}
  \country{USA}
  \postcode{66045}
}

\author{Shawn Keshmiri}
\email{keshmiri@ku.edu}
\affiliation{%
  \institution{University of Kansas}
  \streetaddress{1450 Jayhawk Blvd}
  \city{Lawrence}
  \state{Kansas}
  \country{USA}
  \postcode{66045}
}

\author{Morteza Hashemi}
\email{mhashemi@ku.edu}
\affiliation{%
  \institution{University of Kansas}
  \streetaddress{1450 Jayhawk Blvd}
  \city{Lawrence}
  \state{Kansas}
  \country{USA}
  \postcode{66045}
}





\newcommand{\red}[1]{\textcolor{red}{[#1]}}

\renewcommand{\shortauthors}{Sravan, et al.}
\renewcommand{\shorttitle}{A Vertical Look at UAV Connectivity in the Wild}

\begin{abstract}

Reliable and resilient wireless connectivity is essential for unmanned aerial vehicles (UAVs) and advanced air mobility (AAM) applications in Beyond Visual Line of Sight (BVLOS) scenarios. Terrestrial cellular and emerging Low Earth Orbit (LEO) satellite networks promise enhanced performance for various use-cases. Yet, their comparative performance in the low-altitude airspace with complex $3$D aerial propagation remains largely unexplored. In this paper, we present an open-source measurement platform designed to characterize the performance of commercial cellular (Verizon, a major US provider) and LEO satellite (Starlink) networks through real-world flight tests in rural environments. We implement a comprehensive multi-layer measurement approach spanning physical layer signal metrics, multi-cell network topology, and end-to-end (E2E) application performance. Through an extensive flight campaign with more than $10$ flight tests, $4.5$+ hours of flight time resulting in more than $18$K samples, we present the first detailed, open-source dataset analyzing dual cellular and Starlink performance for low-altitude UAV operations. Our cellular-Starlink comparative results, which are collected \emph{simultaneously at the same time and location}, demonstrate significant performance differences between the two technologies: the LEO
satellite link achieves superior latency performance with $95\%$ of Round-Trip Time (RTT) measurements below $50$ ms compared to $80\%$ under $150$ ms for cellular, and exceptional downlink capacity with $95\%$ exceeding $25$ Mbps versus only $5$ Mbps for cellular. Our analysis on cellular network performance demonstrates that while higher altitudes (e.g., $330+$ m above the sea level) improve signal power by $15-20$ dB via line-of-sight (LOS) propagation, it causes a $3-4$ $\times$ increase in handover rates, which is due to excessive multi-cell visibility rather than signal degradation. Furthermore, we observe asymmetric impacts on the RTT performance due to handovers such that $53.5$\% of handovers improve RTT, but worst-case degradation ($275$ ms) is $2$ $\times$ larger than best-case improvement ($137$ ms). Finally, to generalize our measurement capabilities, we develop a suite of machine learning (ML) models to extend the analysis beyond measured flight paths through spatial and altitude-based prediction of key performance indicators. The measurement platform combined with the ML-based predictions provides a comprehensive toolset for evaluating and optimizing BVLOS UAV connectivity in rural deployment scenarios.

\end{abstract}

\ccsdesc[500]{Networks~Mobile networks}
\ccsdesc[500]{Networks~Network measurement}
\ccsdesc[500]{Networks~Network performance analysis}
\ccsdesc[300]{Networks~Wireless access points, base stations and infrastructure}
\ccsdesc[300]{Computing methodologies~Machine learning}


\settopmatter{printacmref=false}
\renewcommand\footnotetextcopyrightpermission[1]{}

\maketitle

\section{Introduction}
\label{sec:intro}

Unmanned aerial vehicles (UAVs) are rapidly transforming diverse sectors, including logistics, precision agriculture, infrastructure inspection, public safety, and real-time multimedia services~\cite{exp_eval,geraci2022will}. As these applications transition from isolated trials to large-scale deployments, they increasingly rely on reliable connectivity for telemetry, command-and-control (C2), and high-rate payload data. This evolution from standalone to connected UAVs elevates communication from a mere supporting role to a core enabler of safety, reliability, and economic viability. In particular, emerging BVLOS use cases require robust wide-area connectivity capable of supporting three-dimensional ($3$D) mobility, stringent latency and reliability demands, and extended mission durations~\cite{homayouni20223gpp, alsabah20216g, poorvi2025reliable}.

Firstly, the integration of UAVs into existing cellular communication infrastructures has gained significant traction due to their wide-area coverage, mature deployment, and built-in mobility management~\cite{zeng2019accessing,cell_enable,zeng2020uav}. The 3GPP Release $15$ study on enhanced LTE support for aerial vehicles, along with subsequent work in 5G-NR \cite{abdalla2021communications,muruganathan2022overview}, has demonstrated that commercial networks can support UAV C2 and payload traffic. However, several studies (see, for example, \cite{lin2018sky,zeng2019accessing,amorim2018measured}) have revealed significant challenges, including elevated interference, uplink congestion, and coverage variability at higher altitudes. Practical field evaluations further highlight the difficulty of meeting stringent reliability and latency requirements, such as 99.9\% C2 reliability and sub-100 ms one-way delay under complex aerial propagation conditions and dynamic network loading~\cite{homayouni2023verification}. These findings underscore the need for measurement-driven characterization of cellular network performance for aerial platforms and for data-driven models capable of predicting key radio access network (RAN) metrics, such as signal quality and throughput, across spatial and altitude domains. There have been numerous previous works on the topic of cellular-connected UAVs. Particularly, several experimental studies on cellular-connected UAVs consider \emph{mounting a cell phone on the UAV} to measure the performance of the cellular network~\cite{platzgummer2019uav,amorim2018measured,sae2019public,amorim2017pathloss,lin2019mobile}. However, these are limited in scope due to several reasons, such as lack of automatic flight control capabilities, limited radio access measurements, and time synchronization mismatch~\cite{platzgummer2019uav} across different subsystems of the measurement setup.

Secondly, while cellular networks offer strong coverage in many populated and suburban regions, their availability and performance can degrade in \emph{remote, rural, or sparsely populated areas} where many UAV missions, such as environmental monitoring, disaster assessment, and long-range logistics, are most compelling. In these settings, LEO satellite constellations (e.g., Starlink) provide an attractive complementary connectivity option, offering wide-area coverage and relatively low latency compared to traditional geostationary systems~\cite{wang2025exploring, ghafoori2024stars, ghoshal2025replication}. Integrating UAVs with satellite networks can enable resilient connectivity in areas with limited or no terrestrial infrastructure, extend BVLOS mission range, and provide redundancy for safety-critical operations~\cite{baltaci2021survey, zhou2023aerospace, hu2023leo}. There are several prior measurement campaigns to characterize the performance of Starlink (see, for example, \cite{wang2025exploring,qin2024analysis,izhikevich2024democratizing,mohan2024multifaceted,laniewski2024starlink,laniewski2025measuring}) under different mobility patterns such as drive tests. 
However, the performance of satellite links in practical UAV scenarios, which are subject to  antenna pointing constraints, beam misalignment, and fast mobility and maneuvers, remains largely unexplored. 


Finally, relying on a single connectivity substrate, either cellular or satellite, can leave UAV operations vulnerable to coverage holes, capacity limitations, or transient degradations ~\cite{lin20215g, muruganathan2021overview}. Dual connectivity, in which UAVs simultaneously utilize commercial LTE and LEO satellite links, offers a promising path toward resilient and flexible communication architectures, enabling failover, load balancing, and application-aware traffic steering across heterogeneous networks~\cite{vaezi2022evolution}. While a few prior studies have explored the integration of cellular and Starlink networks \cite{hu2023leo,shang2024multi}, these efforts are predominantly restricted to ground-level scenarios. Currently, there is a lack of open, measurement-driven platforms that integrate both commercial cellular and Starlink connectivity on a UAV for simultaneous operation and monitoring. Comparative studies evaluating satellite versus cellular connectivity at matched spatial locations and altitudes remain limited. 

To address the aforementioned research gaps, the contributions of this paper are multi-fold. \textbf{\emph{Firstly}}, we design, develop, and test a comprehensive UAV-based measurement platform with dual cellular and Starlink connectivity. This platform is integrated with our in-house automatic flight system (AFS),  which enables extensive autonomous flight testing campaigns under realistic conditions.   
This platform simultaneously measures both cellular and Starlink networks performance, enabling detailed analysis of their performance and altitude-dependent trade-offs. To the best of our knowledge, no open-source UAV-based measurement platform currently provides synchronized, co-located Cellular and Starlink measurements during BVLOS flight. 
\textbf{\emph{Secondly}}, we develop a comprehensive multi-layer measurement and data collection framework that operates across multiple layers of the protocol stack. At the \textbf{physical layer}, we collect spatiotemporal metrics such as RSRP, RSSI, RSRQ, and SINR, which collectively provide insight into cellular coverage, interference, and link quality as a function of altitude, position, and cell association. At the \textbf{network layer}, we capture information about serving and neighboring cells, handover behavior, and multi-cell network topology.
Furthermore, we collect \textbf{end-to-end (E2E) performance indicators}, including latency, uplink/downlink throughput, and packet delivery statistics, from the UAV platform to our dedicated remote server. 
Our platform enables measuring and correlating all of these metrics for the cellular network and E2E performance metrics for the Starlink network at matched geolocations and altitudes. Our analysis results demonstrate several key findings on altitude-dependent LTE network performance, the relationship between multi-cell visibility and handover behavior as a function of altitude, and performance comparison of LTE vs. Starlink. 
\textbf{\emph{Thirdly}}, to extend our analysis beyond measured trajectories, we leverage ML models to capture complex $3$D propagation patterns and extrapolate performance to unmeasured altitudes. 
Therefore, our empirical performance measurement combined with ML-based prediction algorithms will ultimately provide a comprehensive framework for network performance characterization, coverage assessment, and safe BVLOS route/mission design. In summary, our contributions are as follows: 
\begin{figure}[t]
    \centering
     \includegraphics[width=0.95\textwidth]{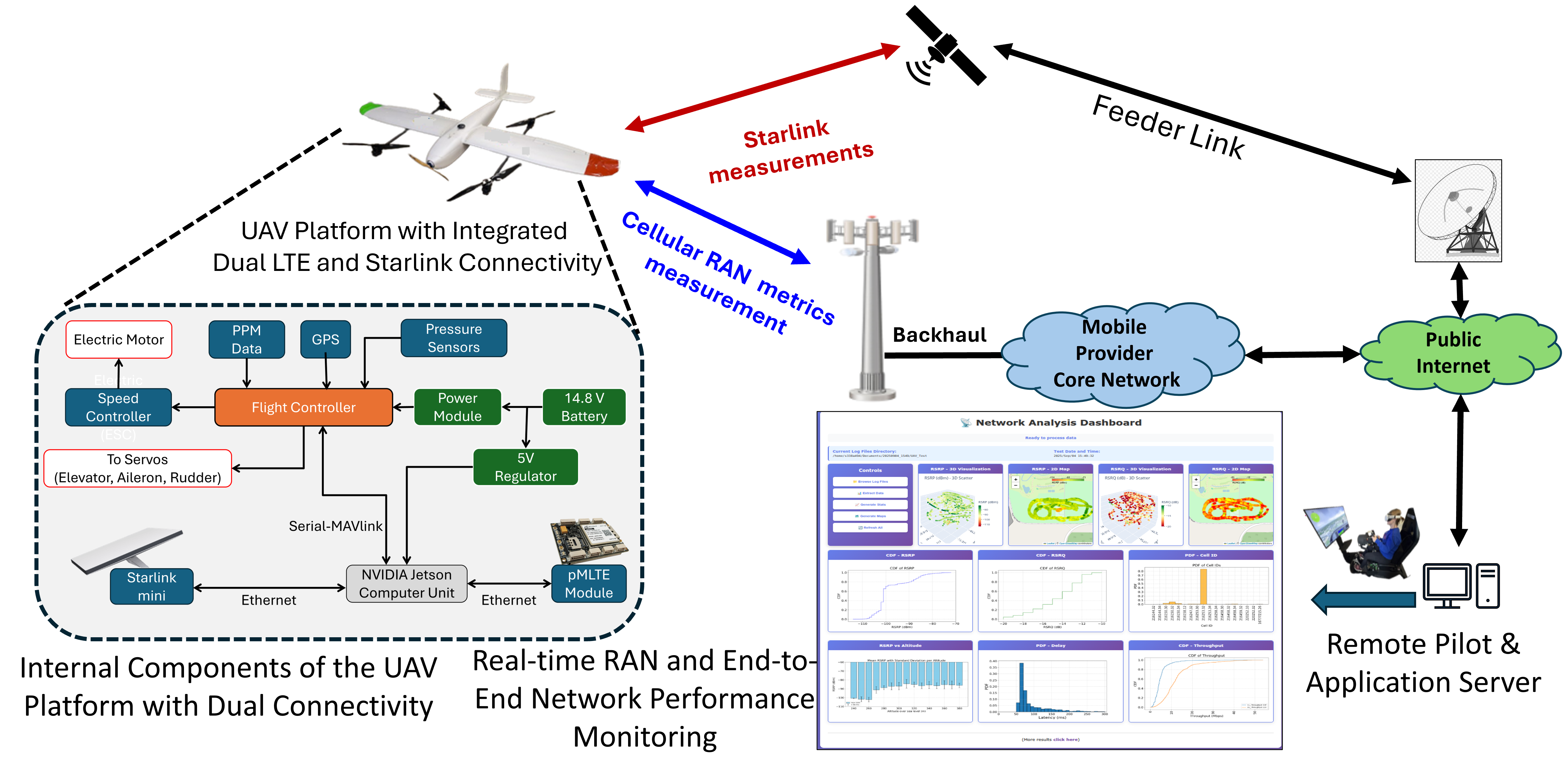}
     \vspace{-.4cm}
    \caption{Overall system model and measurement scenario using UAV(s) that can hover above the ground level equipped with cellular and satellite modems and fly up to an altitude of 400 ft. UAV collects RAN metrics, E2E network measurements, and Starlink measurement data.}
    \label{fig:NSF_UAV}
\end{figure}

\begin{itemize}
[leftmargin = 4mm, itemsep = 0.009in, parsep = 0.005in, topsep = 0.005in]
    \item \textbf{Dual‑connectivity UAV measurement platform}: We design and implement a UAV‑based measurement platform that supports BVLOS operations and integrates both commercial LTE and Starlink connectivity, enabling simultaneous operation and coordinated measurement of heterogeneous wireless networks during real‑world rural flights. To the best of our knowledge, this is the first UAV platform that integrates commercial cellular and Starlink connectivity for simultaneous operation, measurement, and comparative analysis of heterogeneous wireless networks during real‑world flight.
    \item \textbf{Multi‑layer cellular and Starlink measurement dataset}: We develop custom software to capture multi‑layer cellular metrics, including RAN‑level parameters (RSRP, RSRQ, RSSI, SINR) for serving and neighboring cells, network‑level topology information, and E2E performance metrics (latency, uplink, and downlink throughput), alongside Starlink performance measurements collected at the same geolocations and altitudes. This results in a comprehensive dataset collected from May $2024$ to December $2025$, comprising $10$ flight tests with $18,948$ samples. The dataset includes $3.5$ hours of LTE measurements across $7$ flights and over $1$ hour of Starlink E2E measurements across $4$ flights, totaling approximately $4.5$ hours of flight time. For application layer comparison, we have also collected synchronized LTE and Starlink E2E measurements across $2$ flight tests (September $2025$ and December $2025$).
    \item \textbf{Comparative analysis of commercial cellular and Starlink connectivity for UAVs}: By leveraging synchronized measurements at matched spatial locations, we provide a comparative evaluation of cellular and Starlink performance for UAVs, quantifying their respective coverage, latency, and throughput characteristics, and identifying regimes where one technology offers clear advantages or where dual connectivity can provide robustness benefits.
   \item \textbf{ML‑based prediction of key RAN metrics in $3$D space}: To extend the analysis beyond the measured flight paths, we train ML models on the collected dataset to estimate key RAN metrics in both measured and unmeasured spatial and altitude regions, under both random‑split and leave‑one‑altitude‑out validation. These models enable the prediction of network performance in unobserved areas, supporting coverage planning, risk assessment, and route optimization for future UAV missions.
    \item \textbf{Open-source testbed design, software modules, and measurement data:} We provide all platform designs, custom software implementations, and analysis tools as open-source resources to enable reproducible research, facilitate collaborative development, and provide broader access to advanced measurement capabilities for rural network evaluation. To maintain anonymity, references to our open-source repository (which contains accurate geographical information) are omitted. The complete dataset and codebase will be made publicly available in the final manuscript upon acceptance.
\end{itemize}

\noindent 
The rest of this paper is organized as follows. Section \ref{sec:exsetup} presents the platform architecture and measurement methodology, followed by spatial and temporal performance analysis of LTE network in Section \ref{sec:na}. Section \ref{sec:multi-cell} presents our multi-cell measurement results and Section \ref{sec:comparison} provides comparison results between LTE and Starlink networks. In Section \ref{sec:ml} we present our ML-based prediction results followed by concluding remarks and future works in Section \ref{sec:conclusion}.

\section{Platform Architecture and Measurement Methodology}
\label{sec:exsetup}
\begin{wrapfigure}{r}{2in}
    \centering
\includegraphics[width=\linewidth]{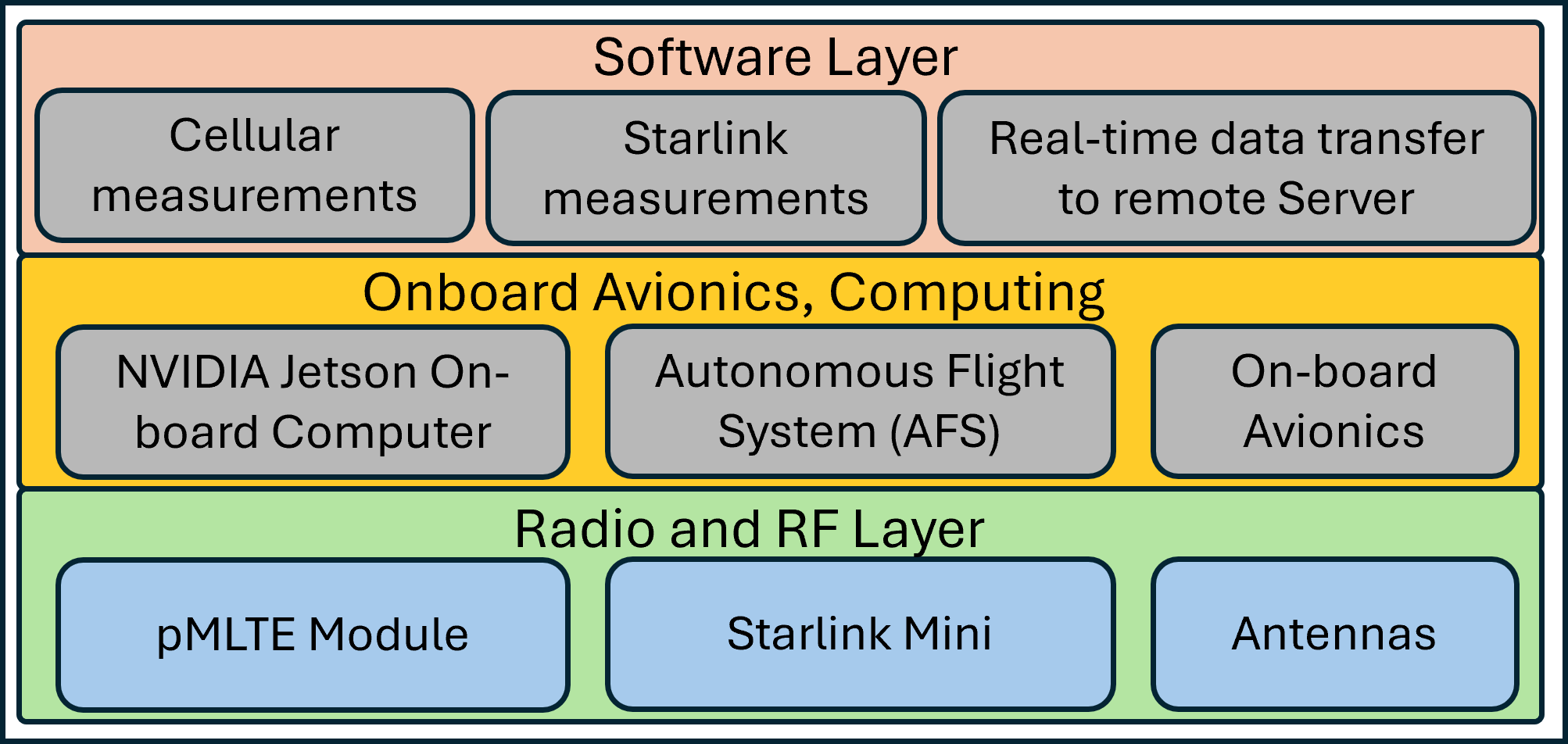}
    \caption{A layered and modular design for integrating dual connectivity technologies with onboard avionics and software logging systems. }
    \label{fig:layered-design}
\end{wrapfigure}

To bridge the gap between theoretical/simulation-based network performance characterization and  real-world scenarios, we develop a modular and open-source measurement platform designed specifically for hybrid terrestrial-satellite network characterization. 
Commercial measurement tools such as Rohde \& Schwarz’s Qualipoc \cite{qualipoc}, TEMS Pocket \cite{TEMS}, and Keysight's NEMO \cite{nemo} (which are used for aerial radio measurements \cite{pathloss_measure,UL_IOI,sae2019public}) or software defined radio (SDR) configurations (e.g., \cite{maeng2023spectrum,volumetric}) are either proprietary (closed-source) or often lack full protocol stack measurements capabilities.

To address this gap, our proposed UAV-based platform integrates automatic flight controller, onboard avionics and compute systems, commercial-grade cellular modem, and Starlink terminal with a synchronized data logging framework.  Figure \ref{fig:layered-design} shows the layered and modular design for our UAV platform, which consists of three distinct layers: \textbf{(i)} radio layer and RF subsystems, \textbf{(ii)} onboard avionics, control, and computing subsystems, and \textbf{(iii)} software modules and data logging subsystems.
In this section, we present the details of our platform architecture and measurement methodology.








\subsection{Platform Design}
\label{sec:platform-design}
\begin{wrapfigure}[10]{r}{1.5in}
\vspace{-0.5cm}
\includegraphics[width=\linewidth, trim=.25cm .5cm 1cm 0cm, clip]{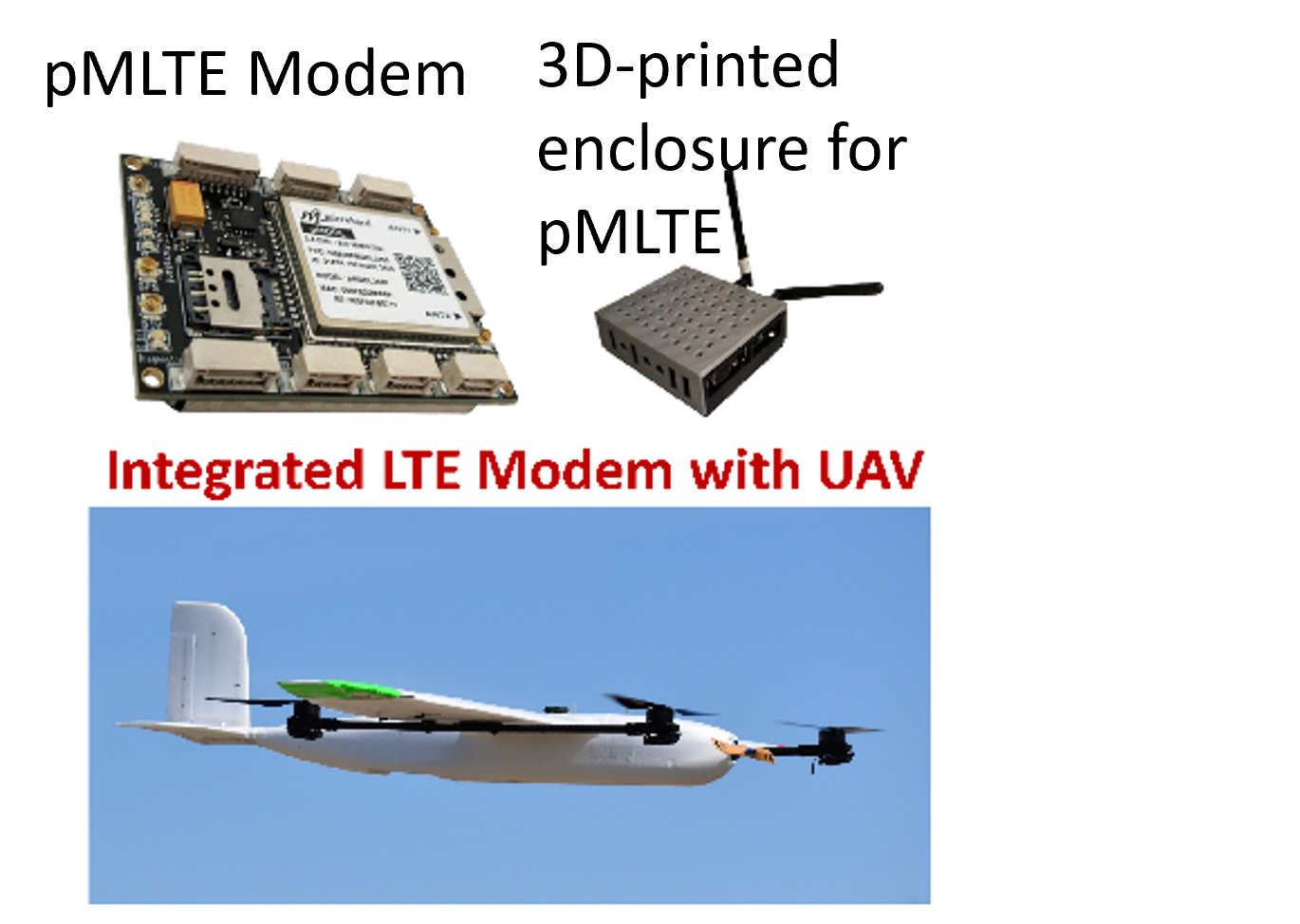}
  \vspace{-0.5cm}
    \caption{UAV platform integrated with LTE modem.}
    \label{fig:LTE-demo}
  
\end{wrapfigure}
\subsubsection{Avionics Subsystems.} For comprehensive and scalable network measurements, 
we designed and assembled large fixed-wing and vertical landing and take-off (VTOL) UAVs ($>12$ lb) (shown in Figures \ref{fig:NSF_UAV}, \ref{fig:LTE-demo}, and \ref{fig:starlink-demo}) for which the avionics suites are fully integrated with commercial LTE modem and Starlink Mini terminal. The VTOL platform combines fixed-wing and rotary-wing (e.g., hexacopters or quadcopters) configurations, which allow the UAV to switch from VTOL mode to fixed-wing mode during cruise flight. This dual-mode capability provides enhanced payload capacity, extended range and endurance, while allowing vertical takeoff and landing operations even from smaller fields. Our UAV platforms are equipped with an in-house Autonomous Flight System (AFS) that features a modular and adaptable architecture built on the Robot Operating System 2 (ROS 2) framework.

The AFS has been successfully integrated into multiple UAV platforms with varying payload capacities and power constraints, enabling flexible mission configurations and scalable measurements.  The AFS is specifically designed for advanced aerospace applications running multiple computationally intensive algorithms in real-time. As shown in Figure~\ref{fig:NSF_UAV}, the onboard avionics centers on the NVIDIA Jetson Orin 16GB~\cite{jetson} as the primary processing unit. 

In Subsection \ref{sec:comm-software}, we present further details on how the Jetson Orin interacts with other components, including the LTE modem and Starlink terminal.

\subsubsection{Radio and RF Subsystem} To characterize the performance of terrestrial and satellite networks, we integrate a commercial-grade cellular modem and a Starlink Mini terminal with our UAV platform. 
\begin{wrapfigure}{r}{1.3in}
    \centering
\includegraphics[width=\linewidth]{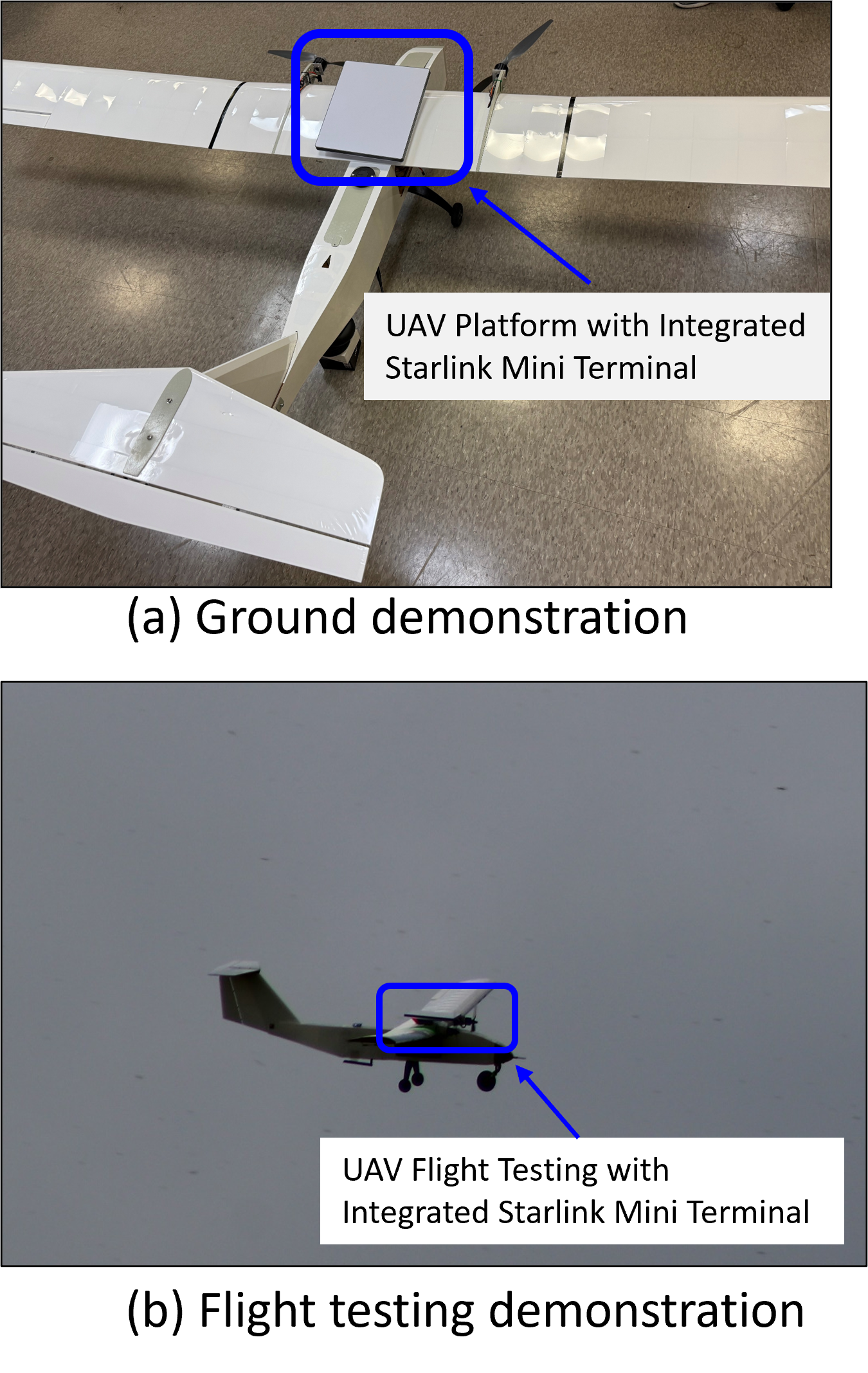}
\vspace{-.75cm}
    \caption{Our UAV platform integrated with a Starlink Mini terminal, shown at ground level and during flight testing.}
    \label{fig:starlink-demo}
\end{wrapfigure}
\vspace{-.1cm}

\textbf{Cellular Modem:} For cellular network measurement, we use the Microhard pMLTE~\cite{pmlte} modem, which provides complete access to serving and neighboring cell RAN parameters through AT command interfaces.  
The cellular modem provides compatibility with globally deployed LTE networks and $3$G/HSPA fallback capability. The module supports LTE FDD with a maximum $150$ Mbps downlink/$50$ Mbps uplink and LTE TDD with a maximum $130$ Mbps downlink/$30$ Mbps uplink. 
Figure \ref{fig:LTE-demo} depicts the pMLTE module, its custom 3D-printed enclosure for airframe integration, and the fully instrumented UAV in flight.


\textbf{Starlink Mini.}
The Starlink Mini is an integrated phased-array satellite terminal combining the antenna subsystem and IEEE $802.11$ac Wi-Fi router in a compact form factor. The terminal utilizes electronically steered beamforming that dynamically adjusts phase relationships to track multiple LEO satellites at approximately $550$ km altitude. 
The satellite interface operates in dual-band frequency allocation: Ku-band with $10.7$-$12.7$ GHz downlink and $14.0$-$14.5$ GHz uplink (primary), and Ka-band at $17.8$-$20.2$ GHz downlink and $27.5$-$30.0$ GHz uplink. 
The system is expected to achieve throughput of $50$ - $100$ Mbps with $25$ - $60$ ms round-trip latency through the LEO constellation to terrestrial gateway infrastructure. 
Due to its lightweight form factor and low power consumption, the terminal is well-suited for UAV platform integration and extended aerial operations. Figure \ref{fig:starlink-demo} shows the integration of the Starlink Mini with the UAV airframe during (a) ground-based configuration and (b) active flight testing (discussed in Subsection \ref{sec:slflight}).

 \textbf{Ground-based Sensitivity Analysis.}
To evaluate the performance of the Starlink Mini terminal for UAV applications, we conducted a ground-based orientation sensitivity analysis. The ground-based analysis evaluated the sensitivity of satellite connectivity to terminal orientation, with the primary objective of identifying potential service degradation scenarios under various spatial configurations that simulate realistic UAV flight altitudes and determining the optimal installation configuration for airborne deployment. Understanding orientation sensitivity is critical for UAV applications, as aircraft undergo pitch, roll, and yaw maneuvers during flight operations that could affect antenna-satellite alignment. The experimental evaluation was performed at an outdoor location with an unobstructed view of the sky. We varied the Starlink Mini's spatial position using a spherical coordinate system. The azimuthal angle $\phi$ represents rotational orientation around the vertical axis, with $\phi = 0^\circ$ aligned with geographic north. Four azimuthal positions were tested: $\phi = 0^\circ$ (north), $\phi = 90^\circ$ (east), $\phi = 180^\circ$ (south), and $\phi = 270^\circ$ (west). The tilt angle $\theta$ simulates UAV pitch and roll maneuvers, with $\theta=0^\circ$ representing the terminal facing directly upward. We evaluated tilt angles of $30^\circ$ and $60^\circ$ relative to the vertical axis.

\begin{figure}[t]
    \centering
    \begin{subfigure}{0.48\linewidth}
        \centering        \includegraphics[width=\linewidth]{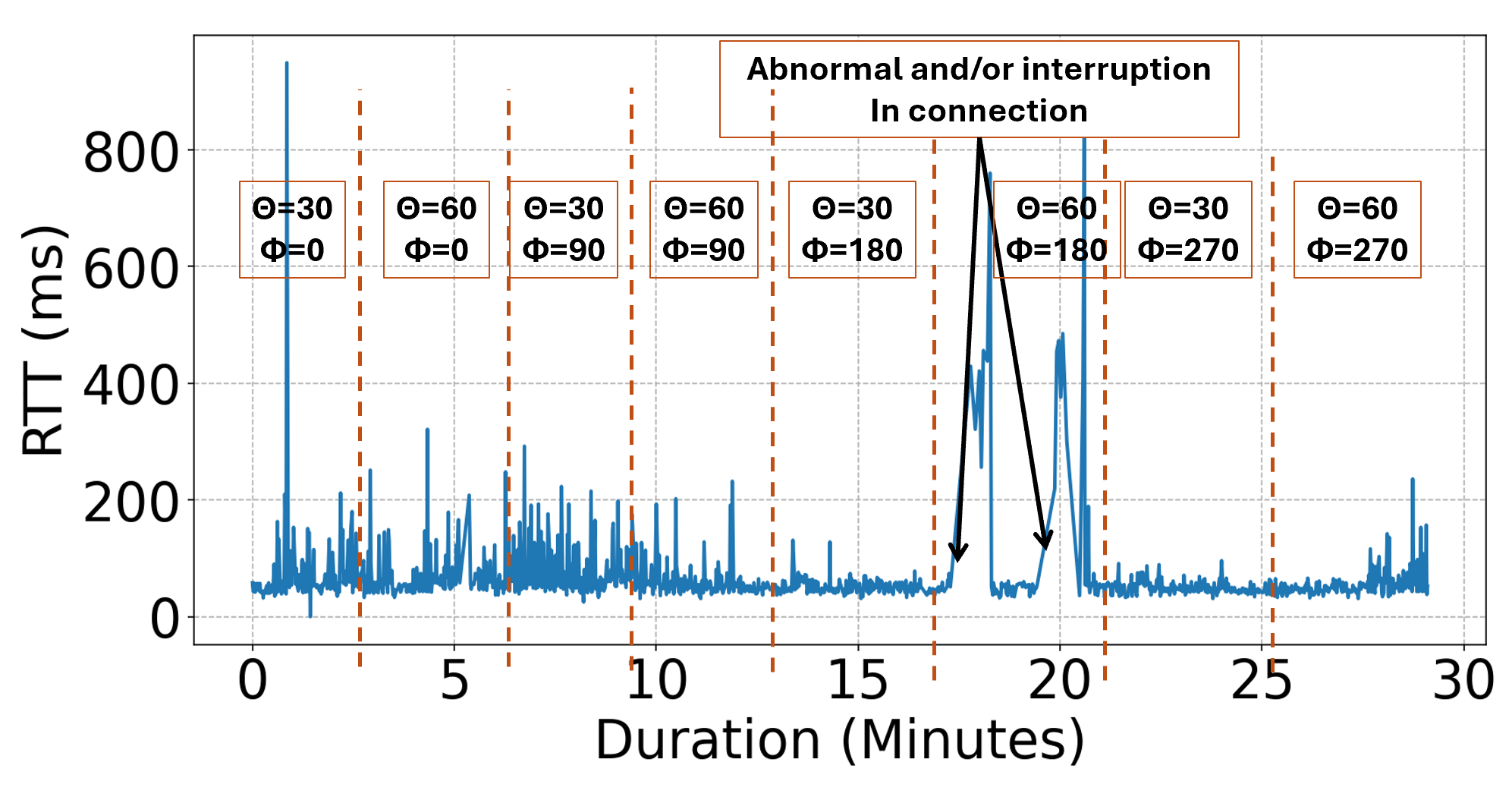}
        \vspace{-0.6cm}
        \caption{RTT.}
        \label{fig:st_rtt_temp}
    \end{subfigure}
    \begin{subfigure}{0.48\linewidth}
        \centering
        \includegraphics[width=\linewidth]{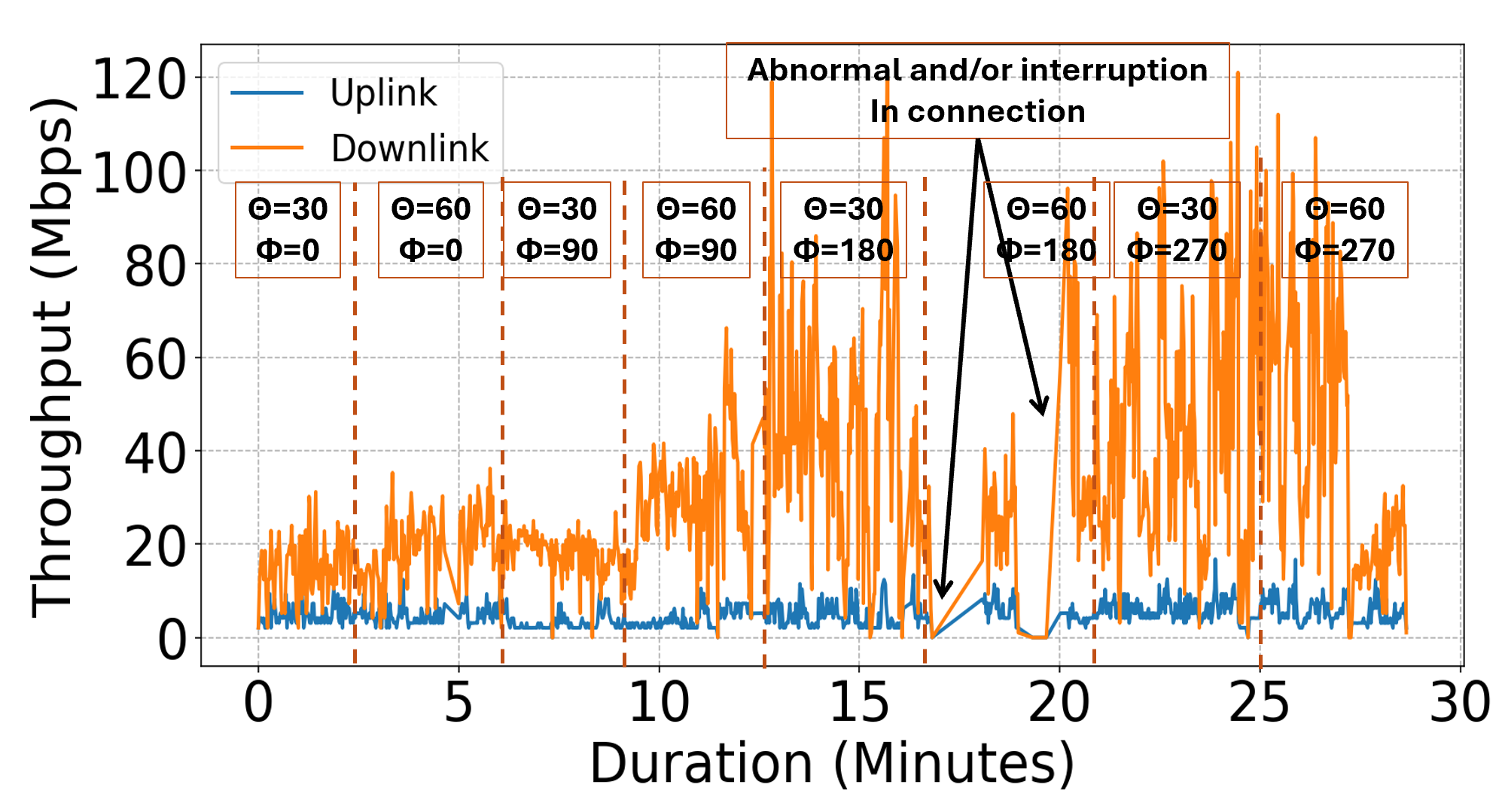}
        \vspace{-0.6cm}
        \caption{Uplink and downlink throughput.}
        \label{fig:st_bidr_temp}
    \end{subfigure}
    \vspace{-0.4cm}

    \caption{Ground-based orientation sensitivity analysis for Starlink end-to-end metrics vs time. }
    \label{fig:ground_test_results}
\end{figure}

The ground-based results, shown in Figure~\ref{fig:ground_test_results}, demonstrate that most tested orientations provided acceptable performance, with RTT values ranging between $40$ ms and $200$ ms, and downlink throughput maintaining $20$-$100$ Mbps. However, critical connection degradation occurred during specific misalignments, particularly at a $60^\circ$ tilt facing south ($\theta = 60^\circ, \phi = 180^\circ$), where RTT exceeded $800$ ms and throughput dropped to near-zero levels due to severe antenna-satellite misalignment. Based on the ground-based findings, the Starlink Mini terminal was mounted on the UAV platform with its surface facing directly upward ($\theta=0^\circ$) to minimize orientation-induced performance degradation during flight maneuvers. 

\subsection{System and Software Architecture} 
\label{sec:software}
\subsubsection{System Architecture.} We implement a comprehensive measurement architecture that coordinates onboard data collection, real-time data transfer, and ground-based analysis framework. As shown in the system architecture in Figure~\ref{fig:NSF_UAV}, a remote application server (e.g., a remote pilot) communicates with the airborne UAV platform. 
In order to support both real-time monitoring and E2E network performance evaluation, the server performs multiple critical functions, including serving as a fixed measurement endpoint for latency and throughput testing, receiving real-time data transmissions during flight operations, and enabling preliminary analysis while measurements are being collected. The fixed endpoint approach eliminates measurement variability introduced by automated server selection in consumer speed testing applications (e.g., Speedtest), enabling consistent comparative analysis across different flight tests and deployment scenarios. The synchronized data collection approach enables various analytical methods, including spatial interpolation to estimate coverage in unmeasured areas, time-series analysis to identify temporal patterns, and correlation studies to understand relationships between position, altitude, and network performance.

\begin{wrapfigure}{r}{2in}
    \includegraphics[width=\linewidth]{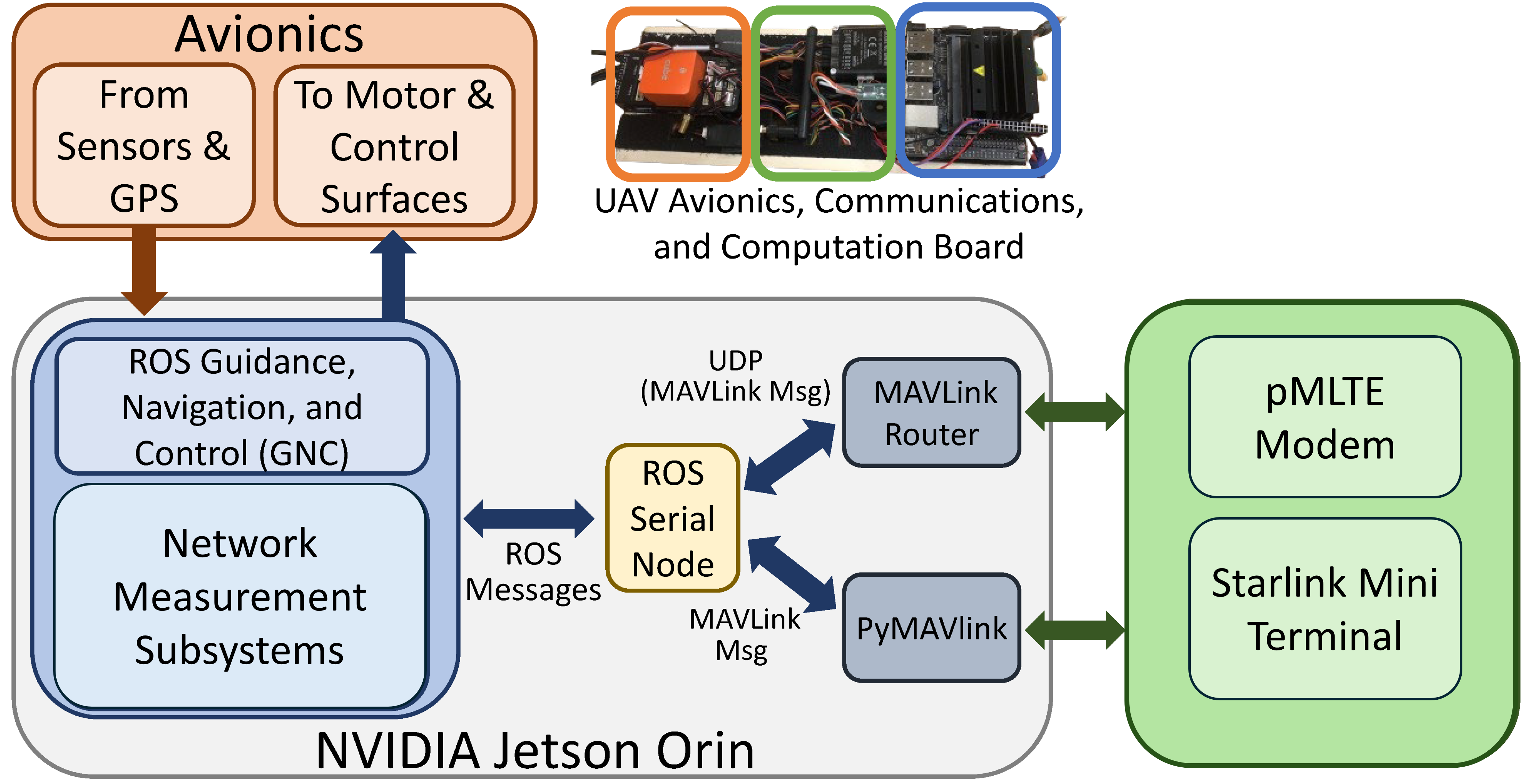}
    \vspace{-.7cm}
    \caption{Internal communication flow of the NVIDIA
Jetson Orin with the pMLTE and Starlink Mini modules.}
\label{fig:internal-jetson}
\end{wrapfigure}
\subsubsection{Communication Software Architecture.}
\label{sec:comm-software}
To define and create data packets, we use MAVLink~\cite{MAVLink}, a ``library for lightweight communication,''  to encapsulate the data being transmitted. MAVLink is an open-source protocol used for communicating with UAVs, and between onboard UAV components (e.g., between flight controller and Orin in Figure \ref{fig:NSF_UAV}). MAVLink is designed to be platform-independent and can be extended with new message types without requiring changes to the current codebase. MAVLink supports a variety of message types, including status, control, and mission messages.
Once the data is formatted into a MAVLink message, it is transmitted via a serial or Ethernet connection to the pMLTE module or Starlink Mini, as shown in Figure \ref{fig:internal-jetson}. The data is then transmitted over the wireless network and received by the remote server. 
In the opposite direction, upon successful arrival of a message, the MAVLink message is captured by onboard Intel's MAVLink Router (IMR). The IMR  scans the incoming data for a valid MAVLink message and once it finds one, it will transmit the packet via UDP to any endpoint specified. The only current endpoint given is a ROS node whose job is to unpack the MAVLink messages and extract the desired information. It then publishes that information to the rest of the ROS system via ROS messages. This ROS node also subscribes to ROS topics that give information about the current location and attitude of the UAV. These subscriptions are spawned in separate threads from the main execution loop of the node and will pull messages into a structure as soon as the data is available. The main execution thread is started in a separate thread within the ROS node and its job is to receive incoming messages and send outgoing messages.

\subsubsection{Graphical User Interface.} To facilitate efficient real-time monitoring and post-flight analysis, we developed a web-based Graphical User Interface (GUI) analysis tool, as shown in Figure~\ref{fig:NSF_UAV}. This tool streamlines the data analysis workflow by providing an integrated platform for log file processing, data extraction, and results visualization. 
The interface includes file browsing functionality that allows users to select measurement log files, data extraction routines that parse the collected metrics, and integrated visualization modules that generate both spatial coverage maps and statistical analysis charts. 
The results are displayed directly within the web interface, providing immediate feedback to the remote pilot during flight operations and enabling rapid iterative analysis during post-flight evaluation. 

\subsection{Multi-Layer Measurement Methodology}
\label{sec:metrics}

\subsubsection{Metrics.} We implement a multi-layer measurement approach capturing network performance metrics across three distinct layers: \textbf{physical layer, network layer, and application layer}. 
We have developed a custom Python-based data logger application deployed on the NVIDIA Jetson to enable automated, high-resolution data collection throughout flight operations. The application operates at a $1$-second sampling interval, providing sufficient resolution to capture signal dynamics during aerial movement. 
At the {physical layer}, the data logger continuously retrieves RAN parameters for both the serving cell and up to three neighboring cells through telnet communication. Collected metrics include RSRP, RSRQ, RSSI, and SINR for each tracked cell. This comprehensive serving and neighboring cell monitoring capability operates continuously without manual intervention throughout flight operations, enabling interference analysis and handover assessment. At the {network layer}, the platform captures network topology identifiers, including Physical Cell ID, Cell ID, and Location Area Code. These identifiers enable spatial coverage mapping and cell-specific performance analysis, distinguishing the platform from single-cell measurement approaches.
At the {application layer}, the platform integrates end-to-end performance evaluation capabilities. The data logger automatically executes Nping~\cite{nping} for packet round-trip time (RTT) measurement and iPerf3~\cite{iperf} for bidirectional throughput testing, both connecting to the dedicated remote server infrastructure. Additionally, packet delivery statistics are collected to assess link reliability. This automated testing framework runs periodically throughout flight operations, capturing latency, throughput, and packet delivery characteristics synchronized with RAN metrics and position data. The fixed remote server endpoint ensures consistent and comparable results across different measurements. Furthermore, we collect all UAV flight information (altitude, direction, speed). All measurement data are captured with GPS coordinates (latitude, longitude, altitude) and high-precision timestamps synchronized via Network Time Protocol, creating a spatiotemporal dataset where every network metric is linked to both physical location and time. 

\subsubsection{Flight Testing and Measurement Area}
\begin{wrapfigure}{r}{2.2in}
\centering
\vspace{-.3cm}
\hspace{-.5cm}
\includegraphics[width=\linewidth]{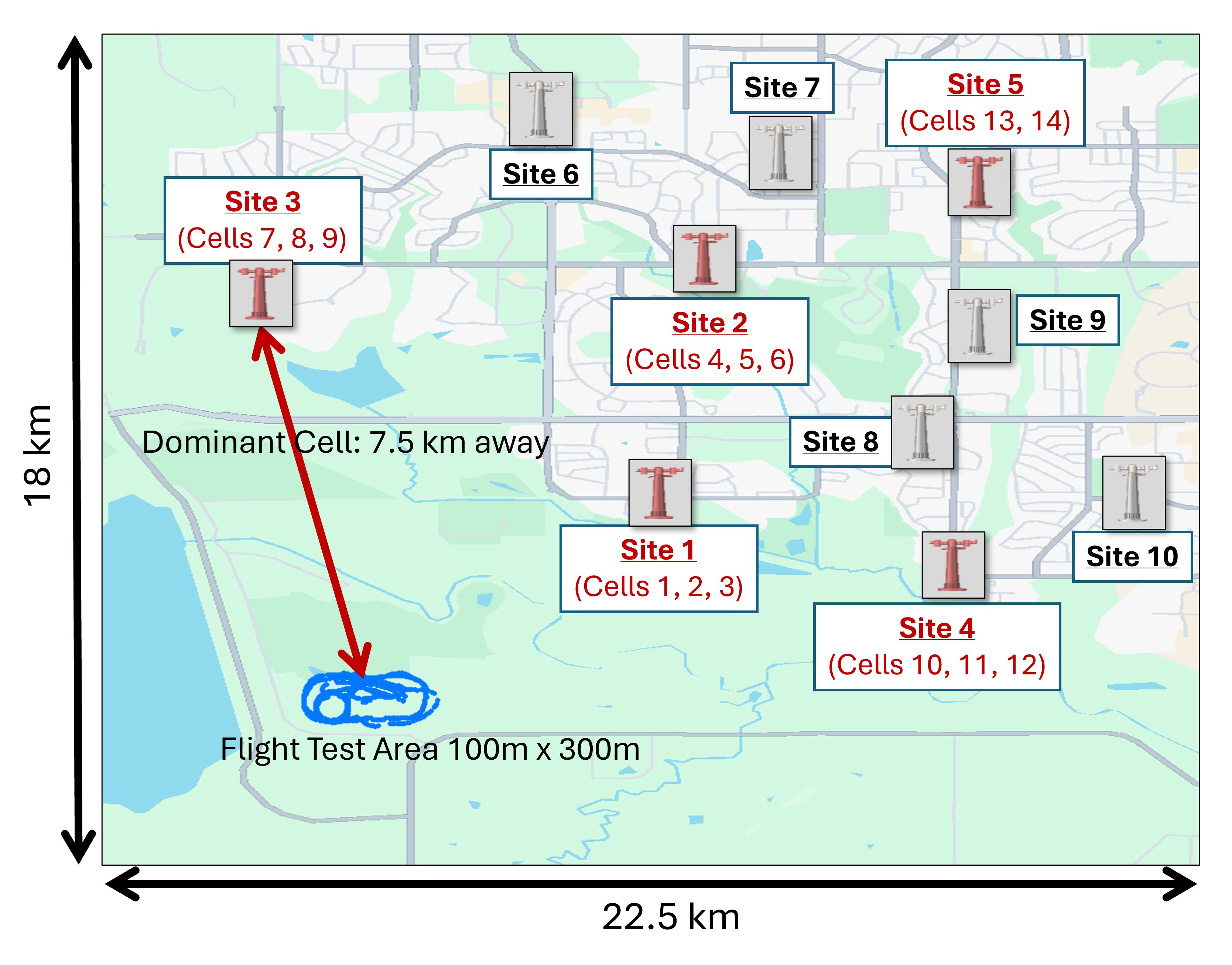}
\vspace{-.5cm}
\caption{Spatial distribution of cell sites around the test area. Each site contains multiple sectors with unique Cell IDs. Red sites contribute to coverage in the test area, while black sites are non-contributing.}
\label{fig:map_CellID}
\end{wrapfigure}
Platform validation was conducted in a semi-rural environment in the U.S., as shown in Figure~\ref{fig:map_CellID}. This test area was selected for its typical rural characteristics including limited cell tower installations with sparse infrastructure deployment and varied terrain with elevation changes affecting propagation. These conditions provide an appropriate environment to study real-world flight operations and validate the platform's three-dimensional measurement capability in rural deployment scenarios. The platform operated on Verizon's commercial LTE network (maximum $20$ MHz channel bandwidth) and Starlink LEO satellite, ensuring measurements reflect real-world operational conditions with realistic interference, load, and network management characteristics absent from controlled testbed environments. Measurements were collected across altitudes from $240$ to $400$ meters above sea level (ASL) (corresponding to $0-160$ meters above the ground level (AGL) in our test area), with data processed using $10$-meter altitude bins to extract altitude-dependent performance trends. This vertical sampling range, combined with horizontal spatial coverage, demonstrates the platform's three-dimensional characterization capability while remaining within Class G airspace regulations for UAV operations ~\cite{kopardekar2016unmanned}. 

Given the presented multi-layer measurement framework, we focus on three key sets of results: 
(1) three-dimensional performance analysis for cellular networks across 
various altitudes (Section~\ref{sec:na}), (2) multi-cell performance characterization via simultaneous tracking of serving and neighboring cells 
performance (Section~\ref{sec:multi-cell}), and (3) comparative end-to-end 
performance evaluation between terrestrial cellular (LTE) and Starlink satellite 
connectivity (Section~\ref{sec:comparison}). 
\section{Temporal and Spatial Characterization of Cellular Network Performance}
\label{sec:na}
In this section, we present our analysis results based on spatiotemporal dataset captured by the platform. First, we quantify the baseline service quality through statistical distribution analysis of key RAN metrics (RSRP, RSRQ, RSSI, and SINR). Second, we investigate three-dimensional performance to characterize altitude-dependent behaviors and empirically validate the trade-off between improved line-of-sight signal (LOS) strength and elevated interference at higher altitudes. 


\begin{figure*}[!t]
    \centering
    \begin{subfigure}{0.48\linewidth}
        \centering        \includegraphics[width=.9\linewidth]{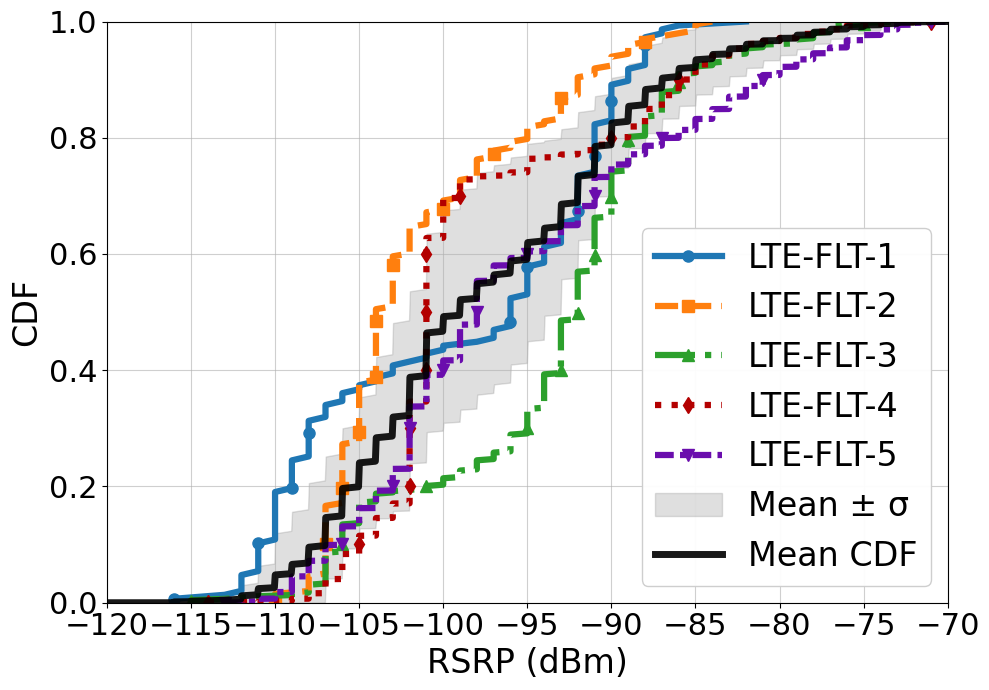}
        \vspace{-.3cm}
        \caption{CDF of RSRP}
        \label{fig:cdf_rsrp}
    \end{subfigure}
    \begin{subfigure}{0.48\linewidth}
        \centering
        \includegraphics[width=.9\linewidth]{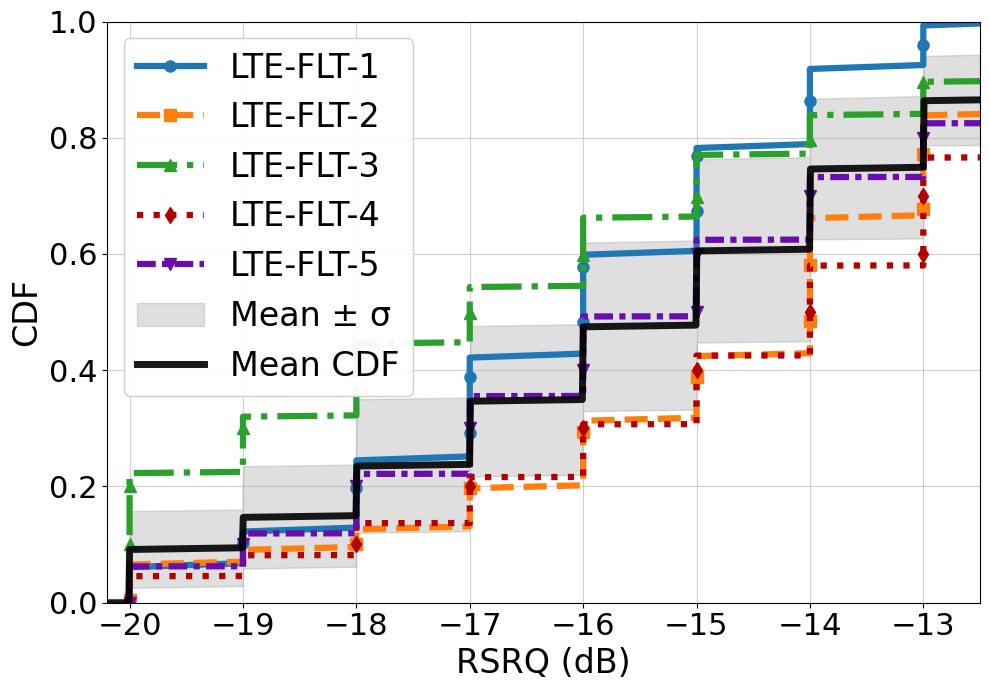}
        \vspace{-.3cm}
        \caption{CDF of RSRQ}
        \label{fig:cdf_rsrq}
    \end{subfigure}


    \begin{subfigure}{0.48\linewidth}
        \centering
        \includegraphics[width=.9\linewidth]{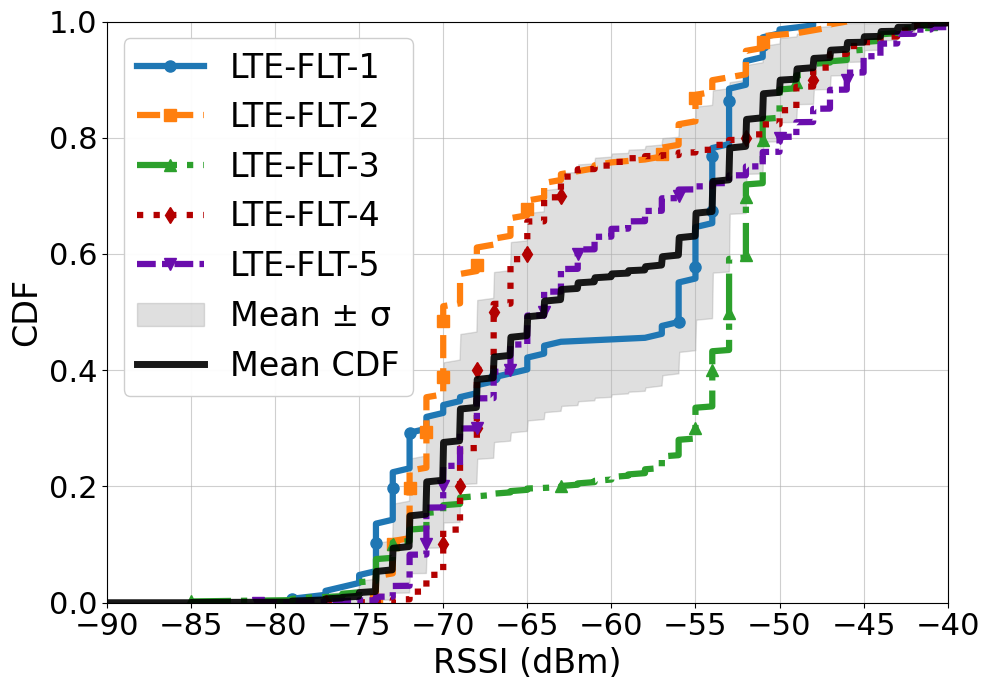}
        \vspace{-.3cm}
        \caption{CDF of RSSI}
        \label{fig:cdf_rssi}
    \end{subfigure}
    \begin{subfigure}{0.48\linewidth}
        \centering
        \includegraphics[width=.9\linewidth]{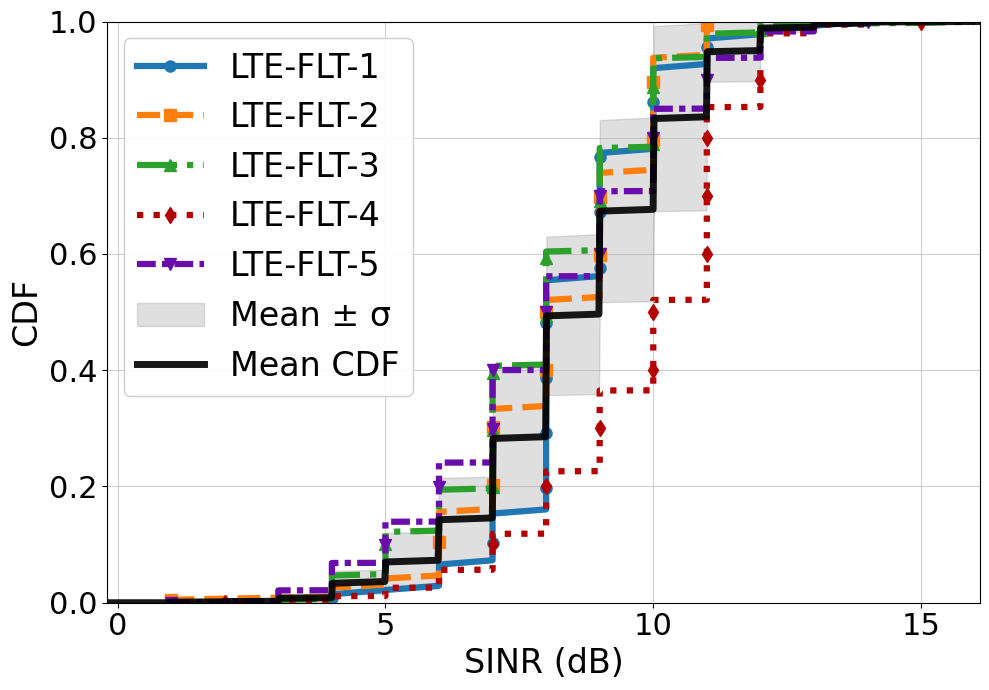}
        \vspace{-.3cm}
        \caption{CDF of SINR}
        \label{fig:cdf_sinr}
    \end{subfigure}
\vspace{-.4cm}
    \caption{CDFs of LTE RAN metrics across five different flight tests denoted by ``LTE-FLT-1'' through ``LTE-FLT-5''. }
    \label{fig:cdf_ran}
\end{figure*}

\subsection{RAN Performance Analysis}
In this part, we use the Cumulative Distribution Function (CDF) to characterize the distribution of RAN performance metrics across the flight testing area. As shown in Figure~\ref{fig:cdf_ran}, CDF analysis was conducted for five flight tests (denoted by ``FLT'') between May 2024 and December 2025, examining RSRP, RSRQ, RSSI, and SINR distributions. The plots present individual flight CDFs alongside mean trends with confidence bands. Since standardized thresholds for quality classification vary across manufacturers and application requirements, we employ the criteria proposed by Teltonika Networks~\cite{teltonika, kpi_thresh} for consistent performance categorization. The following thresholds define poor conditions: RSRP below -$100$ dBm for signal power, RSRQ below -$20$ dB for signal quality, RSSI below -$95$ dBm for channel power, and SINR below $0$ dB for interference immunity.

Figure~\ref{fig:cdf_rsrp} shows RSRP distributions across five flight tests, with the percentage of measurements below the -$100$ dBm threshold varying significantly from $20$\% to $67$\% across different flights. The mean CDF demonstrates consistent coverage trends despite these individual flight variations. Signal quality measurements in Figure~\ref{fig:cdf_rsrq} reveal excellent performance, with $0$\% of measurements falling below the -$20$ dB threshold across all five flights. Channel power conditions in Figure~\ref{fig:cdf_rssi} show $100$\% of measurements well above the -$95$ dBm threshold, indicating uniformly excellent coverage. This high channel power suggests extensive signal overshooting from elevated cell sites. Figure~\ref{fig:cdf_sinr} reveals that $100$\% of measurements exceed the $0$ dB threshold, demonstrating good interference immunity across all flights. The shaded confidence regions indicate measurement variability across the flight tests, with RSRP showing the most variation while RSRQ, RSSI, and SINR remain consistently strong. Overall performance comparison across all four metrics reveals that LTE-FLT-$4$ achieved the best results with the highest RSRQ ($\sim$$-14$ dB) and SINR ($\sim$ $9$ dB) while maintaining competitive RSRP and RSSI values. LTE-FLT-$3$ exhibited the strongest raw signal metrics with the best RSRP ($\sim$ -$93$ dBm) and RSSI ($\sim$ -$56$ dBm), though with slightly lower SINR ($\sim$ $7.5$ dB). 

\begin{figure*}[!t]
    \centering
    \begin{subfigure}{0.48\linewidth}
        \centering
        \includegraphics[width=.8\linewidth]{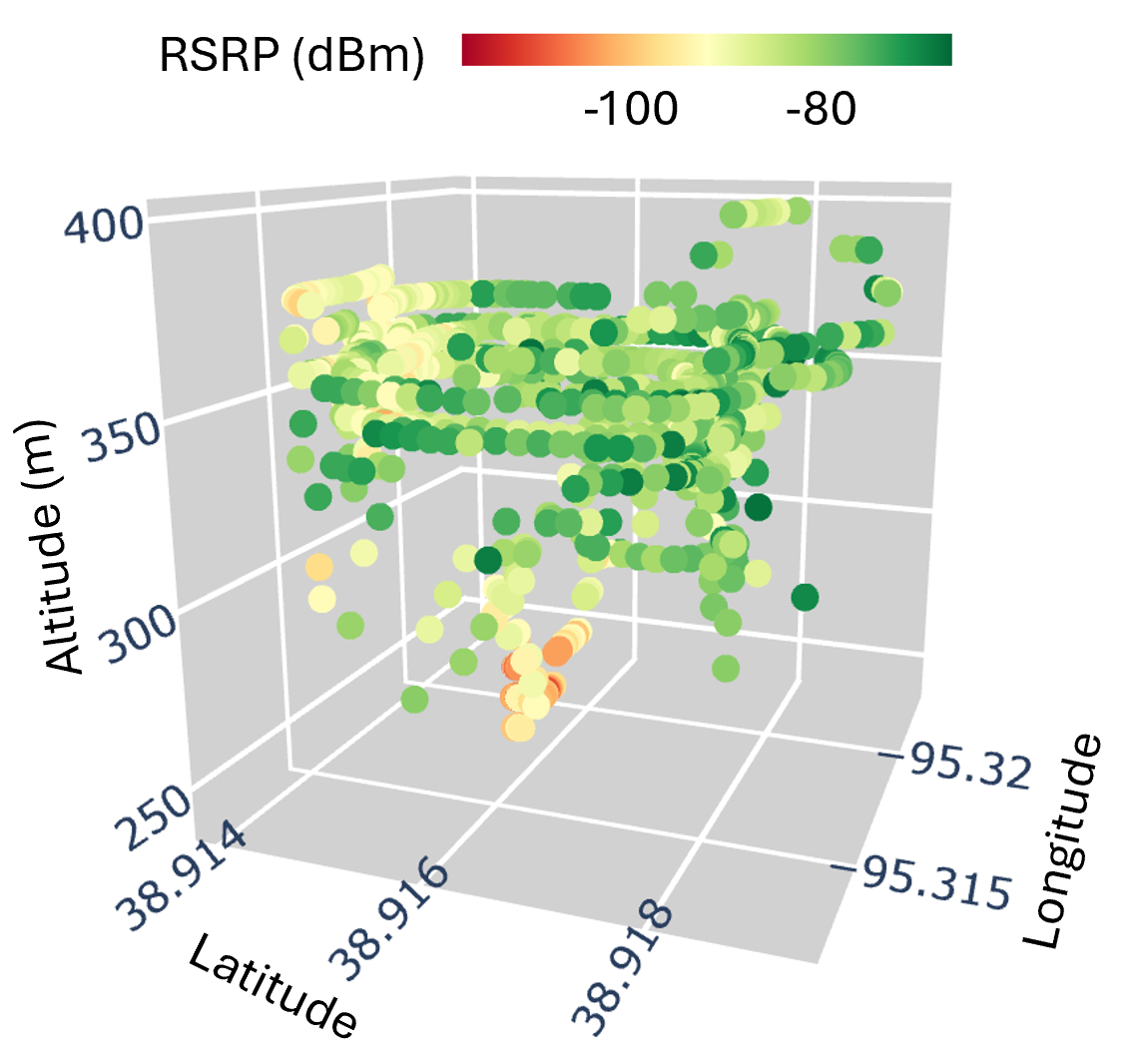}
        \caption{3D RSRP map.}
        \label{fig:3d_rsrp}
    \end{subfigure}
    \begin{subfigure}{0.48\linewidth}
        \centering
        \includegraphics[width=.8\linewidth]{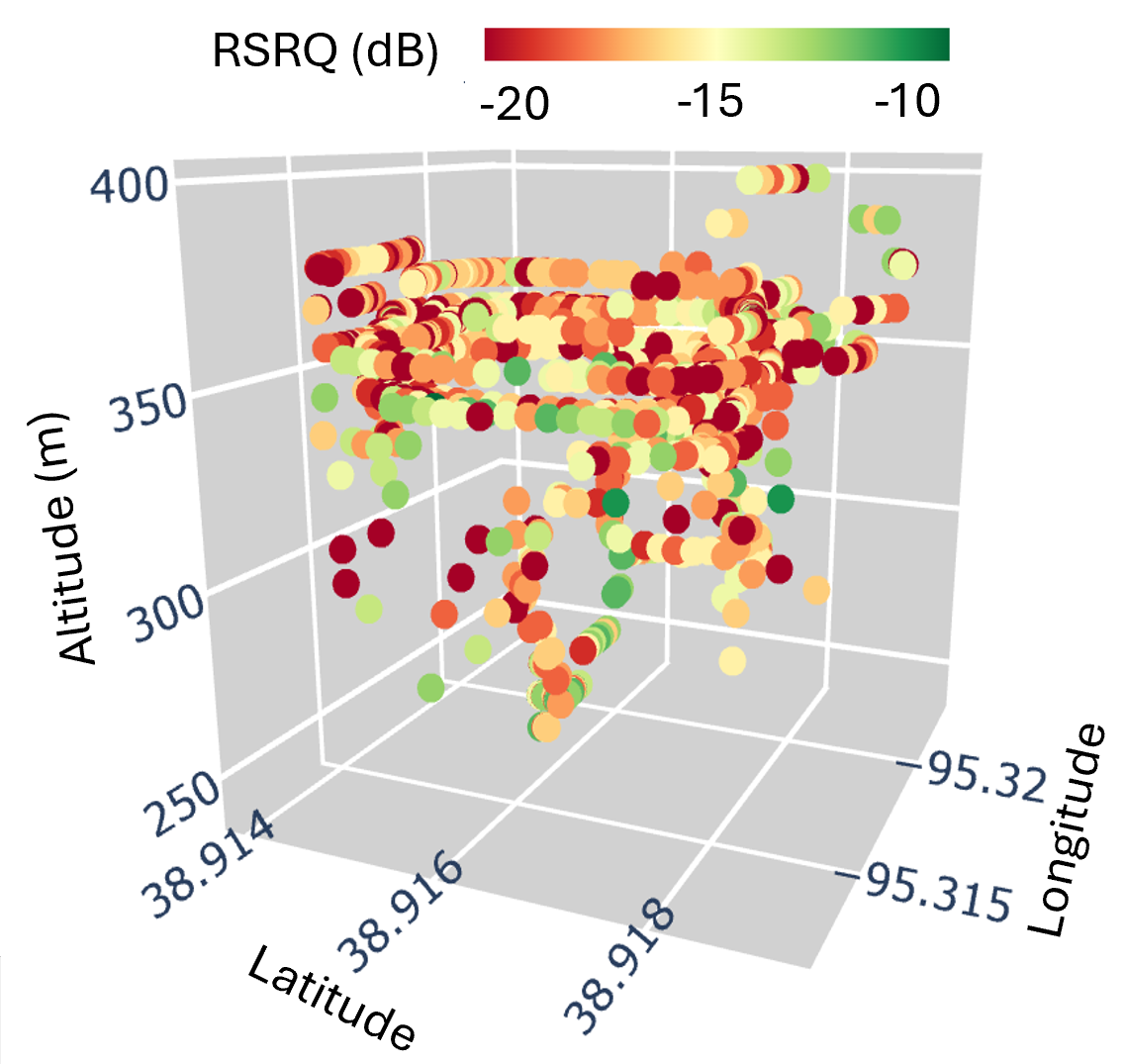}
        \caption{3D RSSI map.}
        \label{fig:3d_rssi}
    \end{subfigure}


    \begin{subfigure}{0.48\linewidth}
        \centering
        \includegraphics[width=.8\linewidth]{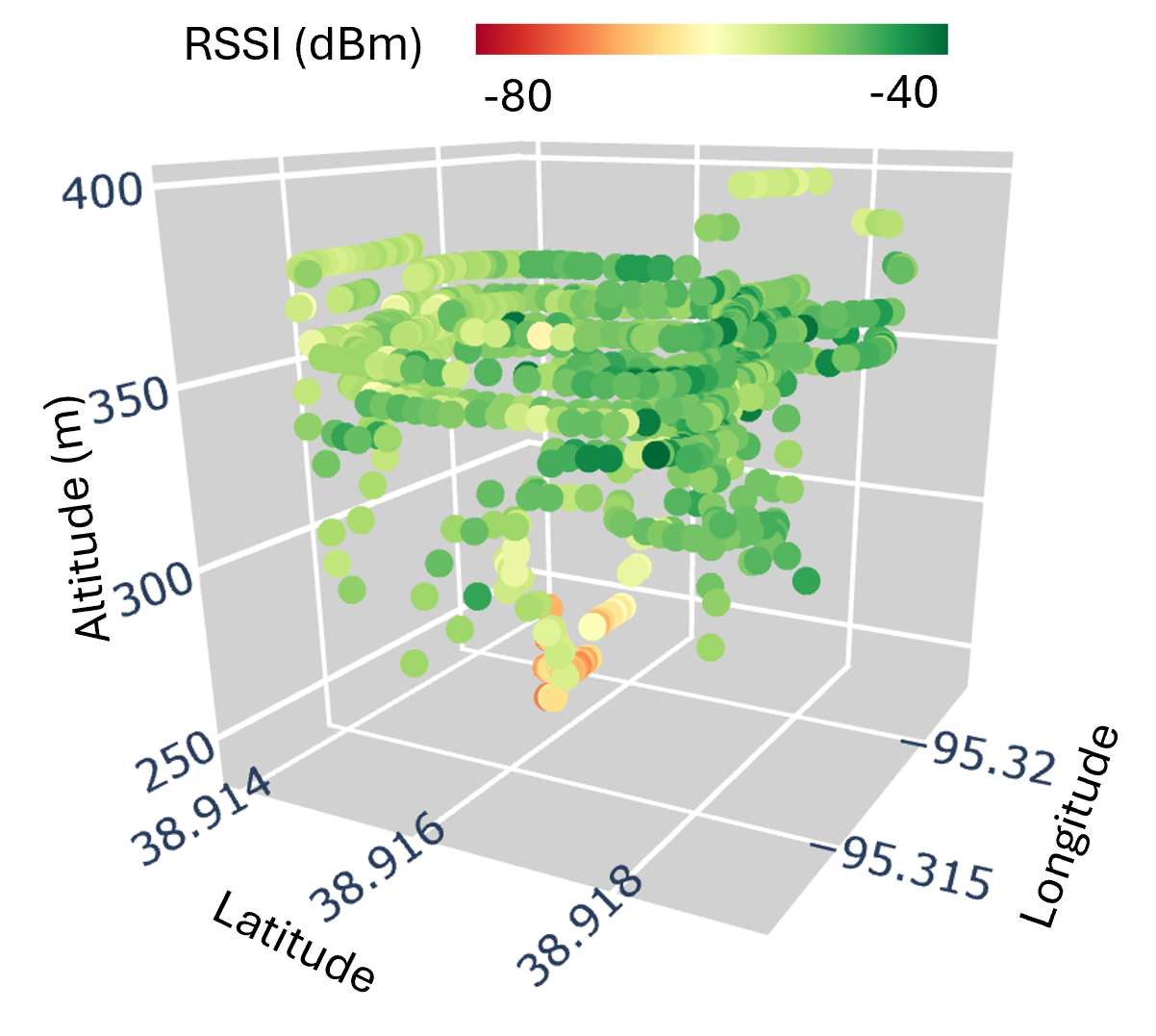}
        \caption{3D RSRQ map.}
        \label{fig:3d_rsrq}
    \end{subfigure}
    \begin{subfigure}{0.48\linewidth}
        \centering
        \includegraphics[width=.8\linewidth]{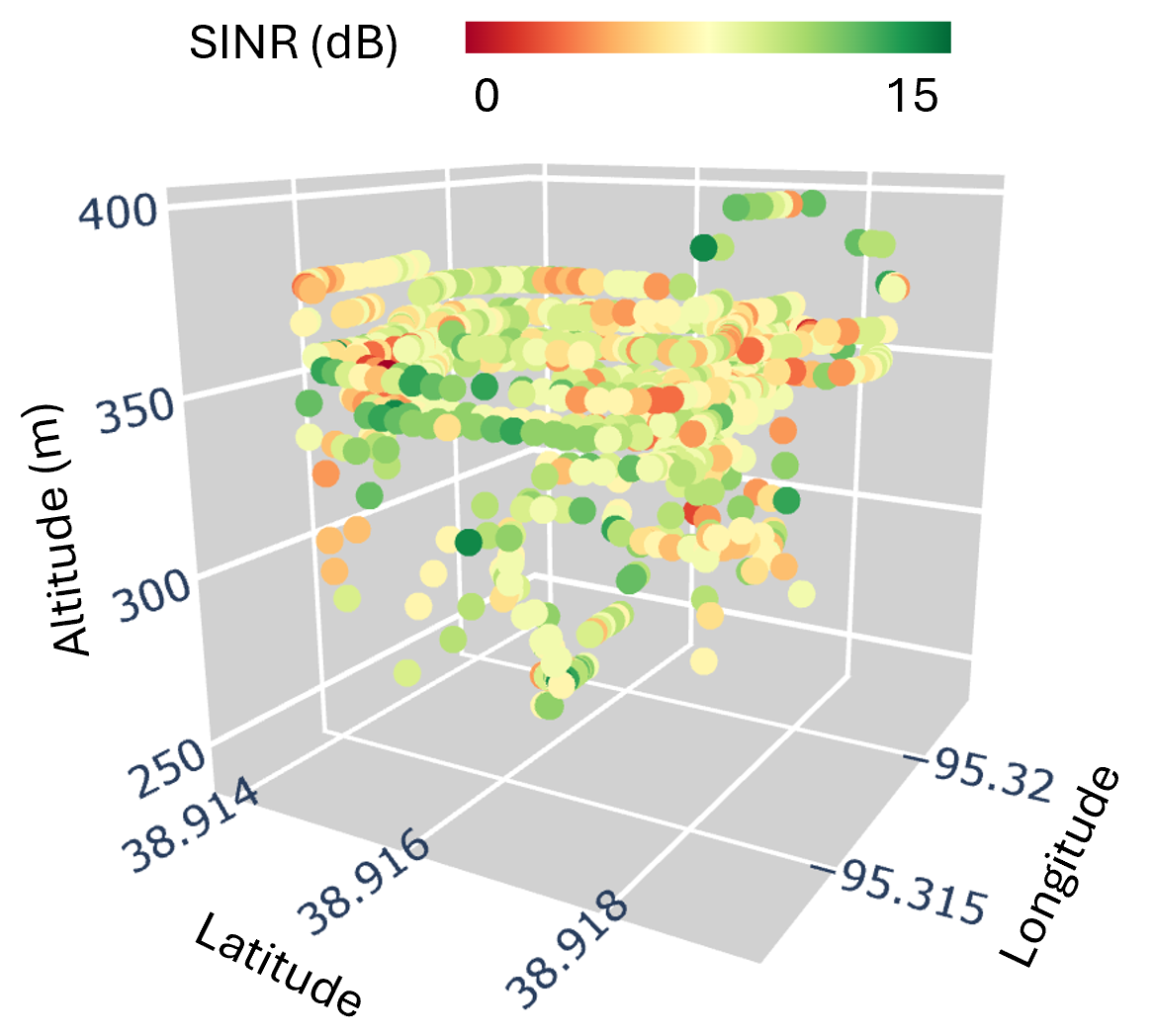}
        \caption{3D SINR map.}
        \label{fig:3d_sinr}
    \end{subfigure}
\vspace{-.4cm}
    \caption{LTE RAN metrics distribution in $3$D maps.}
    \label{fig:3d_metrics_map}
\end{figure*}

\subsection{3D Spatial Performance Analysis}
The platform's ability to collect GPS-synchronized measurements at multiple altitudes 
enables three-dimensional coverage mapping. 
The platform collected measurements from $240$ to $400$ meters ASL, with data processed using $10$-meter altitude bins to extract statistical performance trends. This vertical sampling capability is impossible with ground-based drive testing and provides essential insights for understanding aerial cellular connectivity and coverage prediction for elevated devices. The system generates geospatial visualizations in 3D format, illustrating the 
distribution and behavior of key performance metrics across horizontal position and 
vertical altitude dimensions. Figure~\ref{fig:3d_metrics_map} demonstrates the platform's 3D mapping capability 
through spatial visualization of RSRP, RSSI, RSRQ, and SINR distributions across the 
test environment. The RSRP and RSSI visualizations reveal coverage weak spots at lower 
altitudes corresponding to terrain obstruction, demonstrating acceptable signal strength 
at higher elevations but degradation near ground level in specific regions. The RSRQ 
distribution exhibits spatial variability with improved quality in certain areas, 
indicating varying levels of interference and channel quality across the coverage area. 
The SINR results show fair to poor values with altitude-dependent variations, 
reflecting the complex interplay between signal strength and interference in the rural 
environment.

\begin{figure}[t]
    \centering
    \begin{subfigure}{0.48\linewidth}
        \centering
        \includegraphics[width=0.9\linewidth]{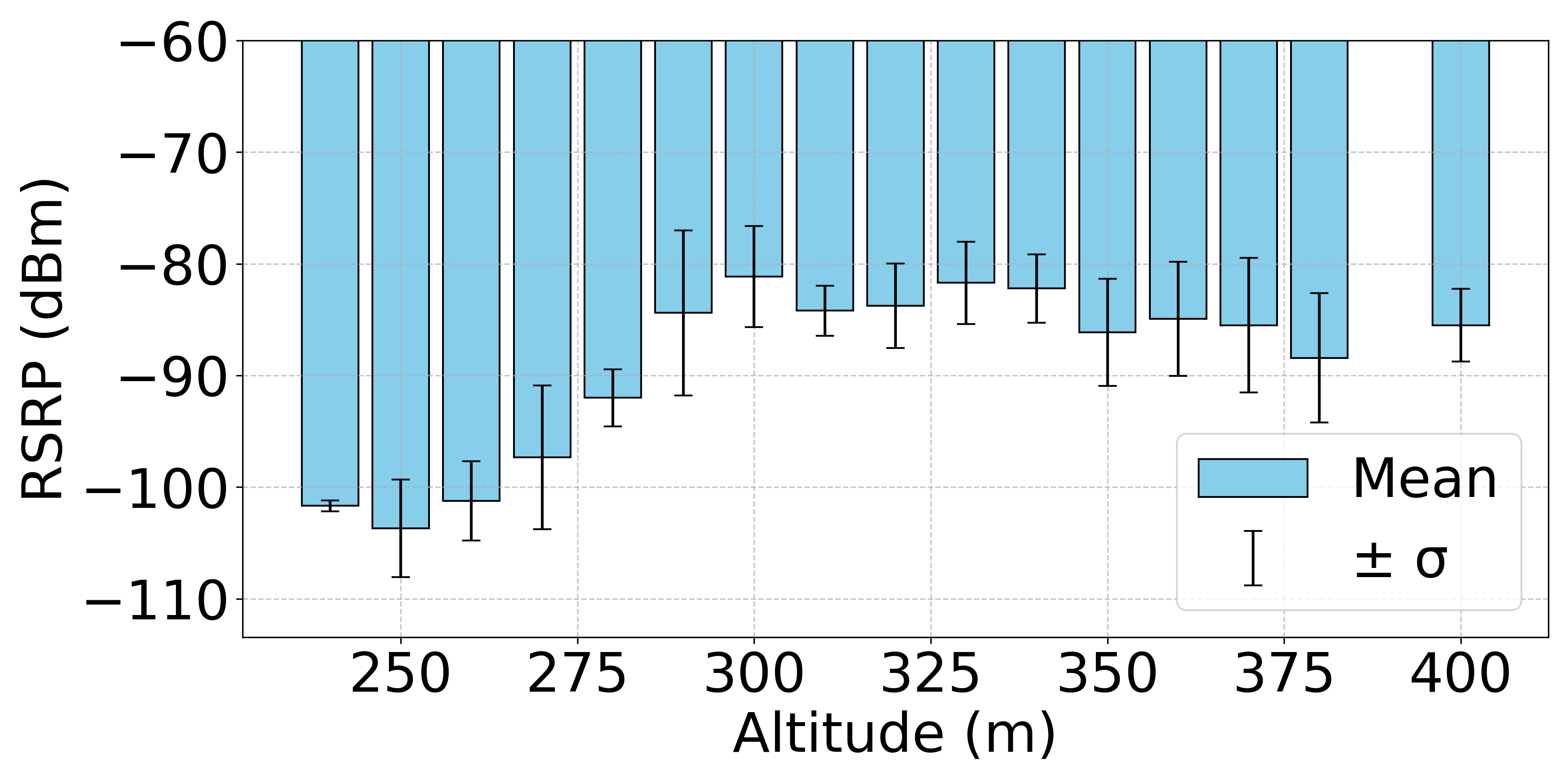}
        \vspace{-.3cm}
        \caption{RSRP vs. altitude.}
        \label{fig:alt_rsrp}
    \end{subfigure}
    \begin{subfigure}{0.48\linewidth}
        \centering
        \includegraphics[width=.9\linewidth]{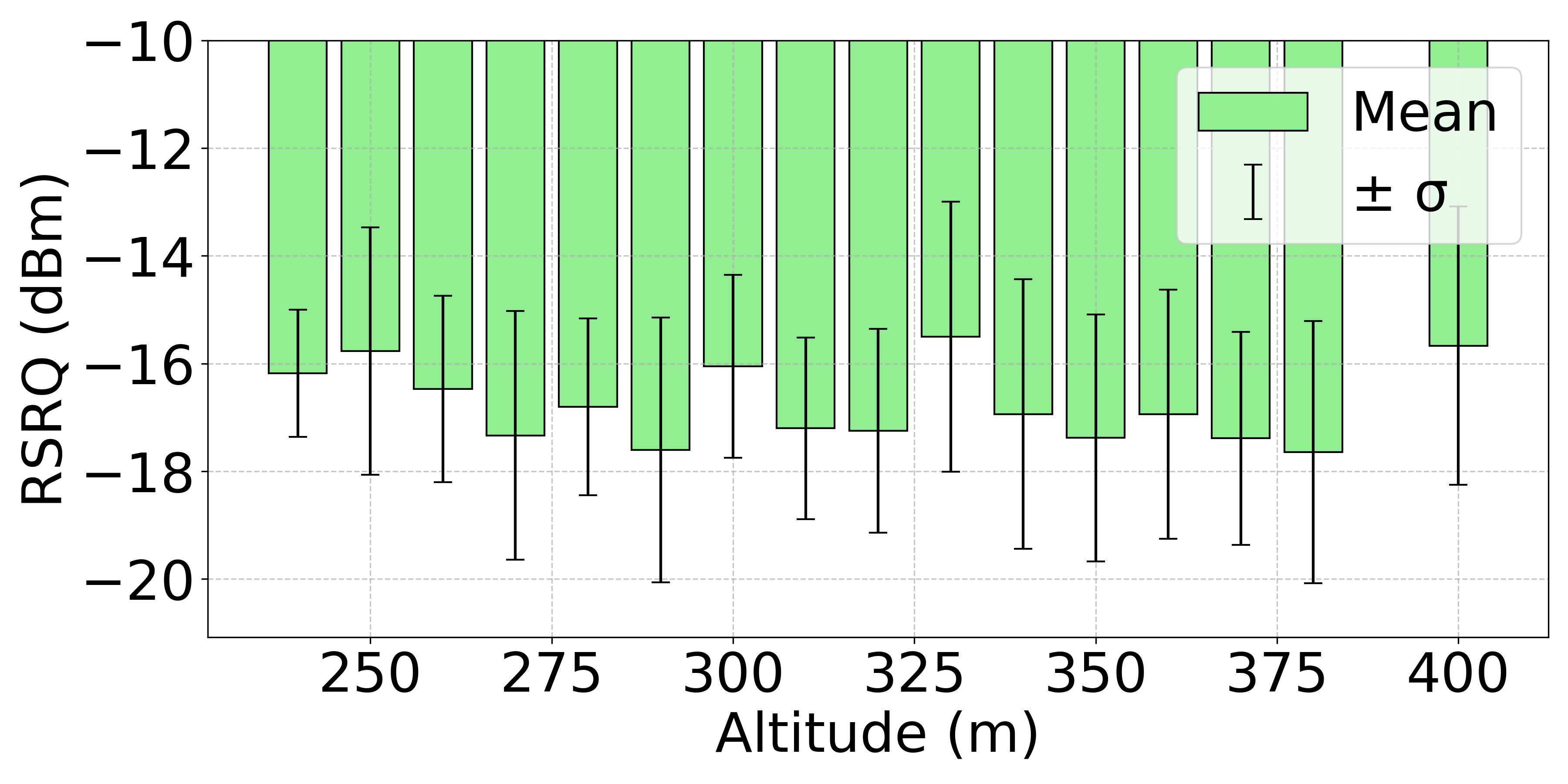}
        \vspace{-.3cm}
        \caption{RSRQ vs. altitude.}
        \label{fig:alt_rsrq}
    \end{subfigure}


    \begin{subfigure}{0.48\linewidth}
        \centering
        \includegraphics[width=.9\linewidth]{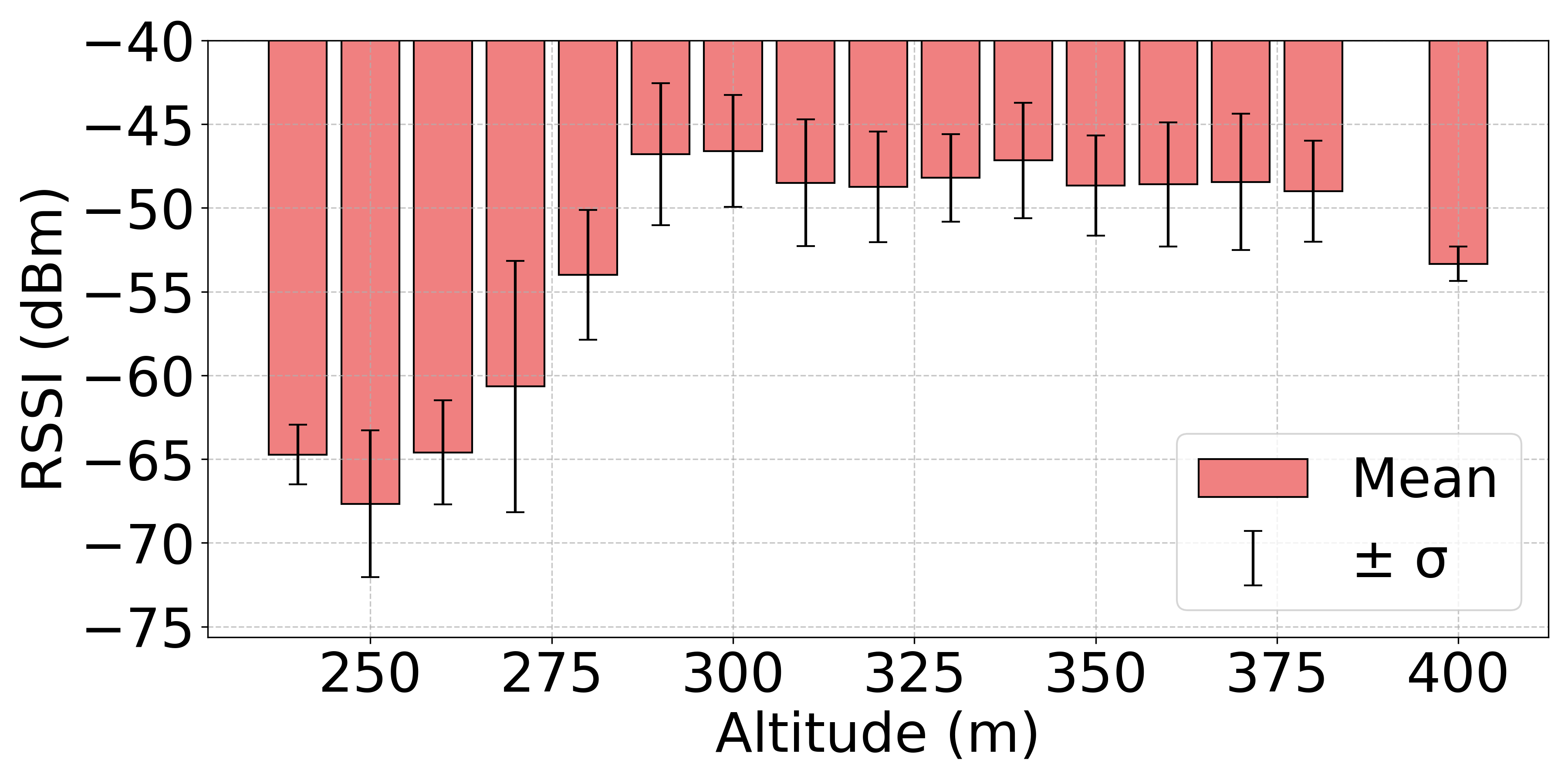}
        \vspace{-.3cm}
        \caption{RSSI vs. altitude.}
        \label{fig:alt_rssi}
    \end{subfigure}
    \begin{subfigure}{0.48\linewidth}
        \centering
        \includegraphics[width=.9\linewidth]{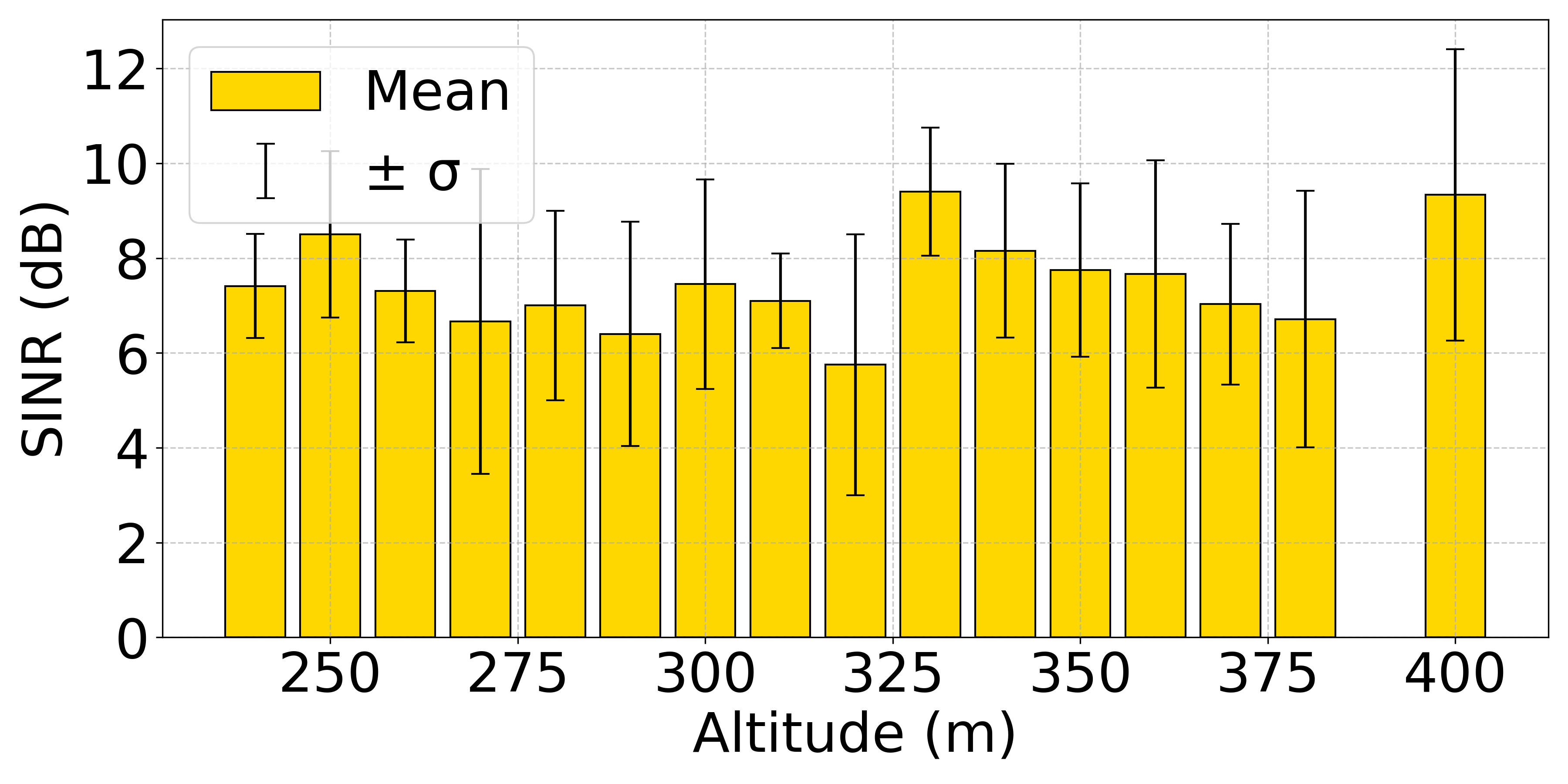}
        \vspace{-.3cm}
        \caption{SINR vs. altitude.}
        \label{fig:alt_sinr}
    \end{subfigure}
\vspace{-.4cm}
    \caption{LTE RAN metrics as a function of altitude. As altitude increases, the RSRP values improve due to enhanced LOS conditions; however, this leads to increased interference from neighboring cells.}
    \label{fig:ran_alt}
    \vspace{-.4cm}
\end{figure}

Furthermore, Figure~\ref{fig:ran_alt} demonstrates the platform's ability to reveal altitude-dependent signal strength trends. The RSRP measurements show progressive improvement with increasing altitude, attributed to enhanced LOS propagation conditions between the UAV and the base stations. At lower altitudes, terrestrial obstacles, including terrain features and vegetation, cause significant signal attenuation. As the platform ascends, the probability of achieving direct LOS increases, reducing path loss. This characterization capability enables network operators to understand vertical coverage profiles and predict performance for aerial users. The platform's multi-altitude measurement capability reveals a critical trade-off for aerial cellular connectivity: while signal power improves with altitude, signal quality degrades due to increased multi-cell interference. At elevated positions, the UAV gains unobstructed visibility to multiple base stations simultaneously, including cells beyond the intended serving cell. This fundamental altitude-dependent behavior, where improved LOS conditions simultaneously enhance desired signal reception and interference from neighboring cells, can only be characterized through vertical measurements that the platform enables. These results provide comprehensive insights for 
identifying coverage gaps, interference patterns, and altitude-dependent signal quality variations in three-dimensional propagation environments.

    

        

 \section{Multi-Cell Performance Analysis}
 \label{sec:multi-cell}
Aggregate signal strength is a primary indicator of connectivity quality in cellular networks. However, reliable BVLOS operations rely fundamentally on network redundancy and the ability to execute seamless handovers during flight time. In this section, we extend our analysis from the serving cell's isolation to a broader multi-cell perspective. We first investigate the spatial distribution of serving cells in Section~\ref{sec:dominance}, and demonstrate that it is possible a ``dominant'' cell could provide the majority of aerial coverage despite its suboptimal radio performance. Next, in Section \ref{sec:handover}, we quantify the network performance by evaluating the availability and signal quality of neighboring cells to assess handover readiness. We present a detailed analysis on the handover behavior as a function of RAN metrics, flight altitude, and impacts on RTT performance.

\begin{figure*}[t]
    \centering
    \begin{subfigure}{0.48\linewidth}
        \centering
        \includegraphics[width=.9\linewidth]{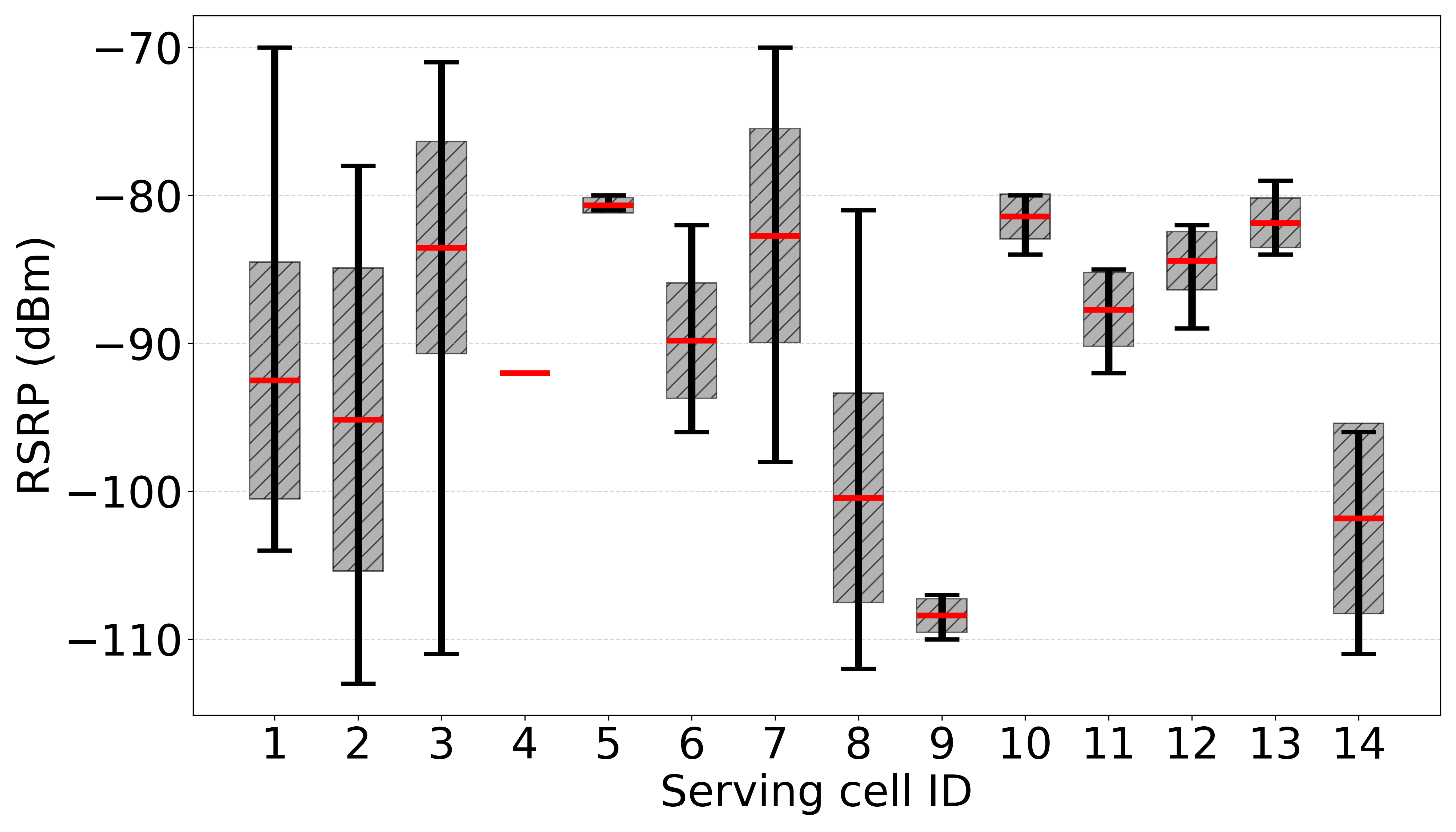}
        \vspace{-.3cm}
        \caption{Serving cells RSRP.}
        \label{fig:stat_rsrp}
    \end{subfigure} 
    \begin{subfigure}{0.48\linewidth}
        \centering
        \includegraphics[width=.9\linewidth]{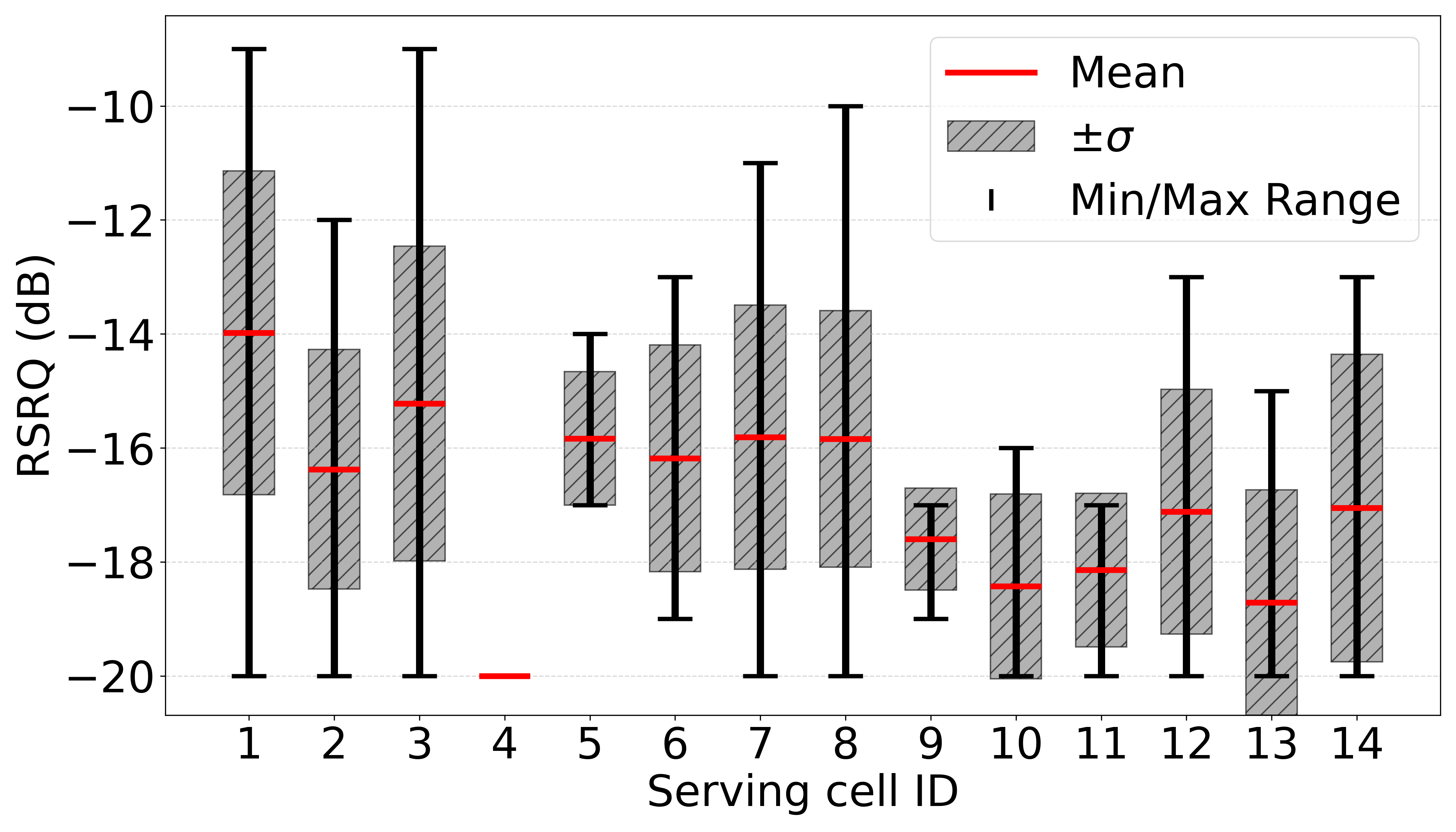}
        \vspace{-.3cm}
        \caption{Serving cells RSRQ.}
        \label{fig:stat_rsrq}
    \end{subfigure}


    \begin{subfigure}{0.48\linewidth}
        \centering
        \includegraphics[width=.9\linewidth]{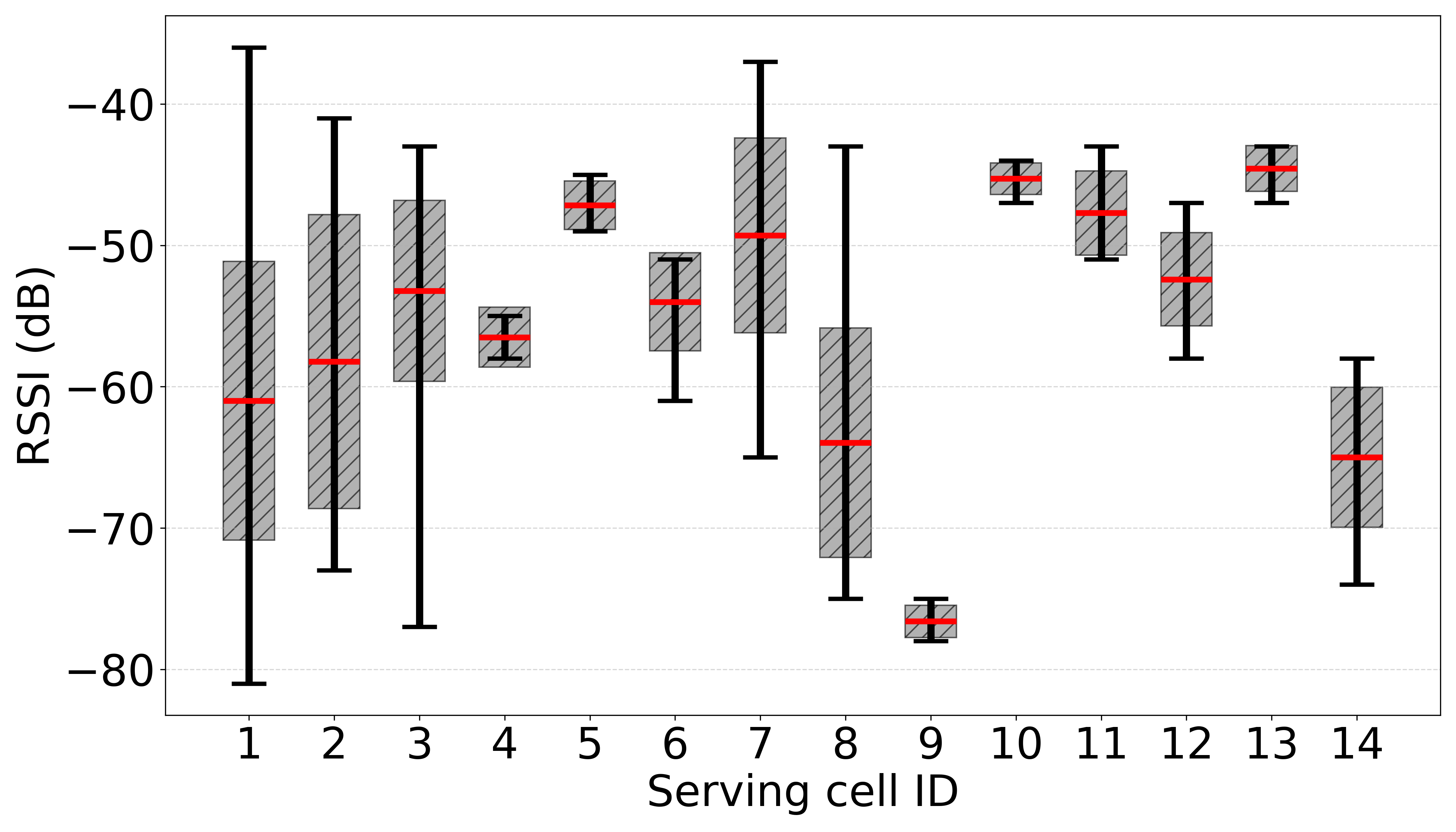}
        \vspace{-.3cm}
        \caption{Serving cells RSSI.}
        \label{fig:stat_rssi}
    \end{subfigure}
    \begin{subfigure}{0.48\linewidth}
        \centering
        \includegraphics[width=.9\linewidth]{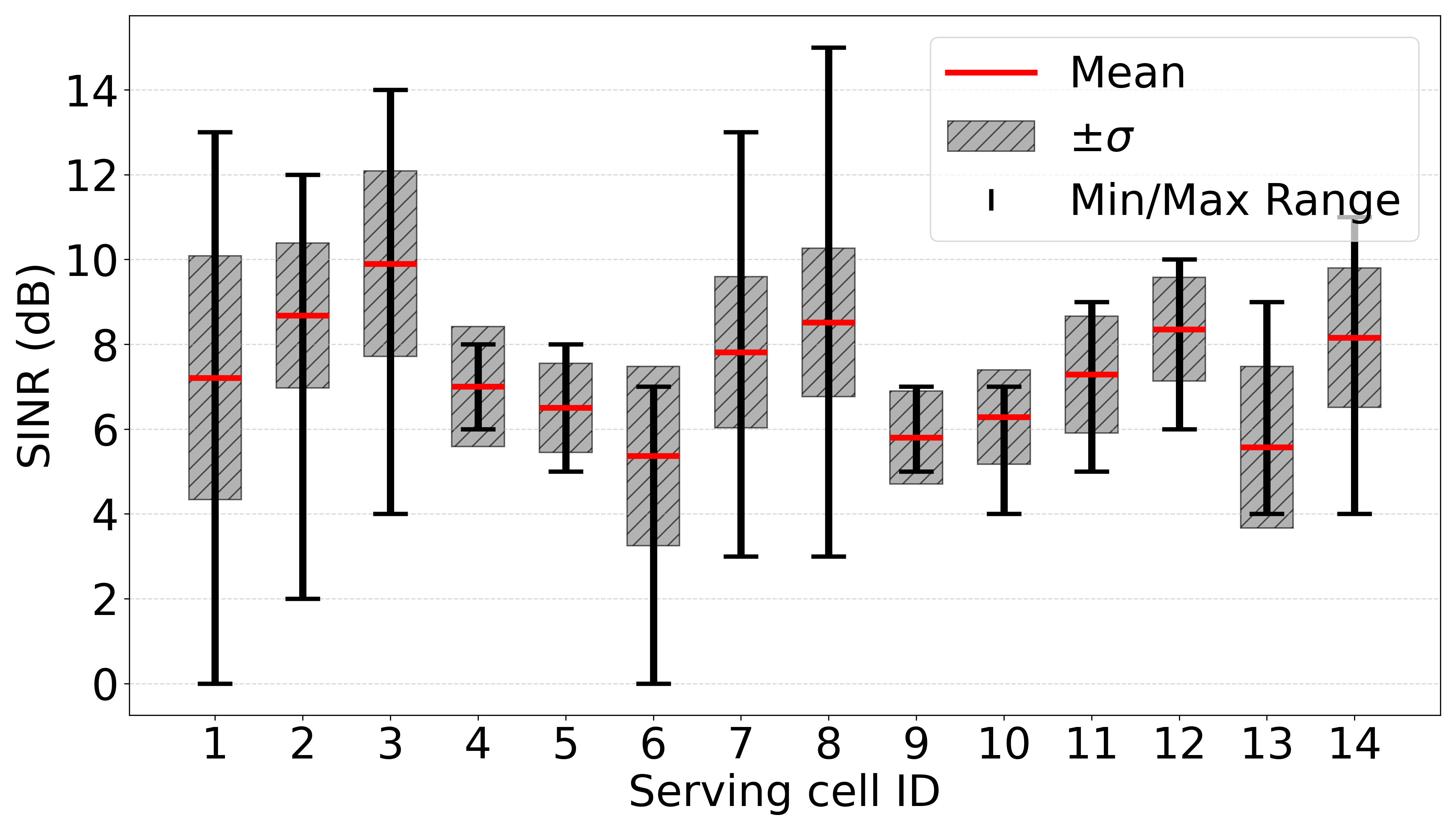}
        \vspace{-.3cm}
        \caption{Serving cells SINR.}
        \label{fig:stat_sinr}
    \end{subfigure}
\vspace{-0.2cm}
    \caption{LTE RAN metrics for different serving cells. In total, signals from 14 distinct cells were received in our testing area shown in Figure \ref{fig:map_CellID}.}
    \label{fig:stat_ran}
\end{figure*}
\subsection{Multi-Cell Coverage Dominance and Spatial Distribution Analysis}
\label{sec:dominance}

\begin{wrapfigure}{r}{2in}
\centering
\includegraphics[width=\linewidth]{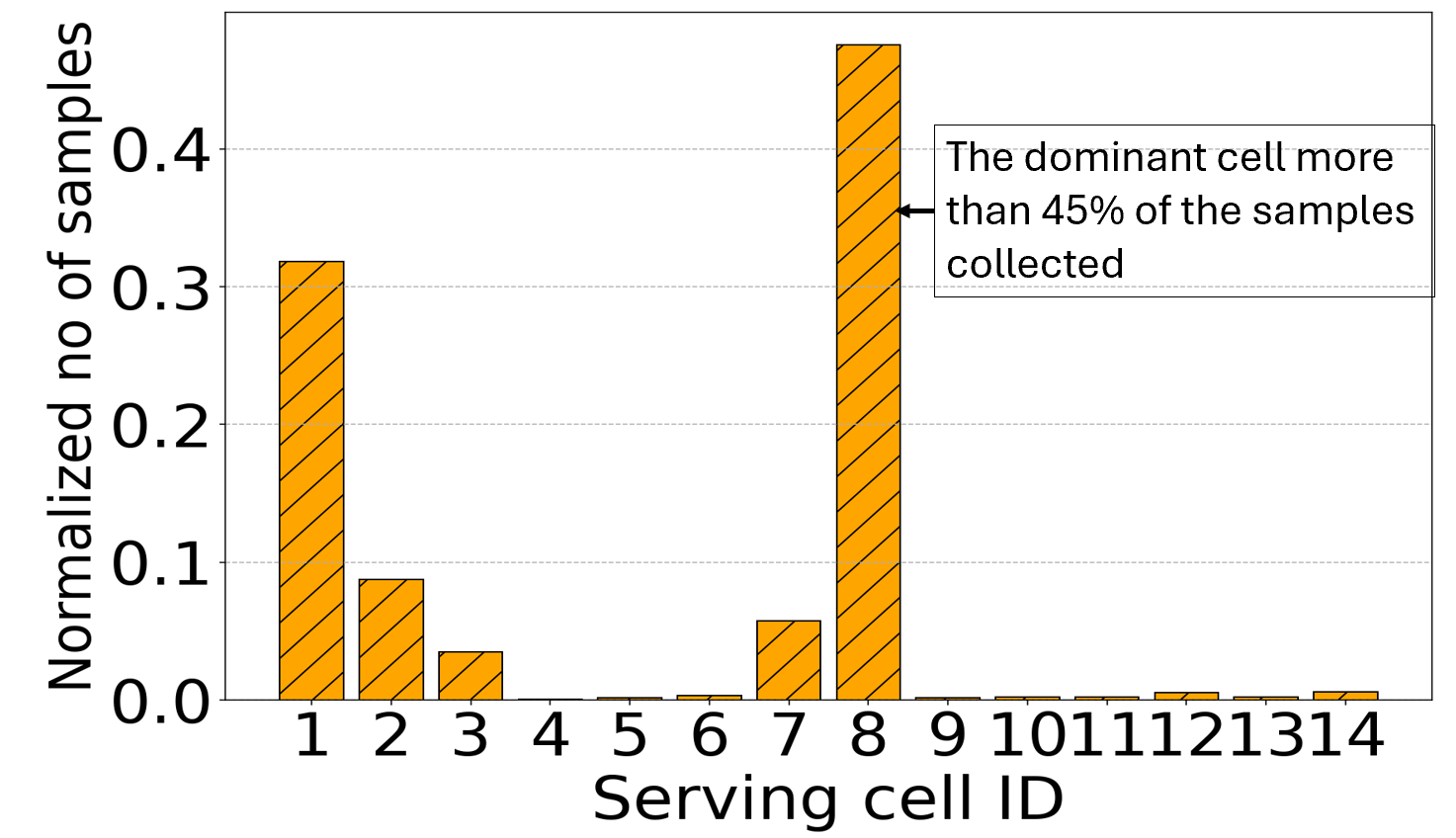}
\vspace{-.8cm}
\caption{Normalized number of samples received from each cell.} 
\label{fig:PDF_CellID}
\centering
\end{wrapfigure}

While  typical RAN metrics such as RSRP, RSRQ, RSSI, and SINR indicate signal quality, 
they alone cannot determine network performance without considering each cell's spatial coverage contribution. A cell may exhibit excellent radio metrics but serve only a small fraction of the area, minimizing its impact on user experience. Conversely, a cell with 
suboptimal parameters that dominates the coverage area significantly affects overall service quality. To investigate these behaviors, we capture the normalized Cell ID occurrence frequency across all measurement samples collected from the measurement area (Figure \ref{fig:map_CellID}). Results in Figure~\ref{fig:PDF_CellID} show that Cell $8$ dominates the coverage area  with approximately $45\%$ of measurement samples, followed by Cell $1$ with approximately $30\%$, and Cell $2$ with approximately $10\%$. The remaining cells (Cells $3$, $7$, $13$ and $14$) each contribute less than $5\%$. 
However, we observe interesting behaviors when contrasting this coverage dominance with the RAN performance metrics shown in Figure~\ref{fig:stat_ran}.

In particular, Cell $8$, serving approximately $45\%$ of the test area, exhibits poor signal strength with mean RSRP around $-100$ dBm and mean RSSI around $-65$ dBm, but shows moderate RSRQ ($\sim -16$ dB) and fair SINR ($\sim 8$ dB). Cell $1$, accounting for $30\%$ of samples, demonstrates moderate RSRP ($\sim -93$ dBm) but maintains good RSRQ ($\sim -14$ dB) and SINR ($\sim 7$ dB). Together, these two cells serve approximately $75\%$ of the test area that directly defines the network quality experienced by the majority of aerial users in the area. Cell $2$, contributing $10\%$ of coverage, exhibits good performance across all metrics (RSRP $\sim -97$ dBm, RSRQ $\sim -16$ dB, SINR $\sim 9$ dB) but due to limited spatial presence, does not contribute to overall service quality. The remaining cells, despite their individual metric variations, have minimal practical impact due to serving less than $5\%$ of the coverage area each.

Spatial analysis using CellMapper~\cite{cellmapper}, shown in Figure~\ref{fig:map_CellID}, confirms that most detected cells are appropriately positioned for test area coverage. Cells $13$ and $14$, installed on high-elevation terrain (confirmed via Google Earth analysis), reach the test area through signal overshooting despite contributing less than $5$\% coverage and exhibiting poor signal strength (average RSRP $\sim$-$108$ dBm). The site corresponding to Cells $1$, $2$, and $3$ shows variable coverage contribution influenced by antenna orientation patterns, as verified through CellMapper directional data. For real-world rural UAV applications in this deployment area, network optimization efforts must prioritize Cells $8$ and $1$, as improvements to their performance yield the greatest impact on overall service quality by addressing the $75\%$ of the coverage area they collectively serve.

\begin{figure}[!t]
    \centering
    \begin{subfigure}{0.48\linewidth}
        \centering
        \includegraphics[width=.9\linewidth]{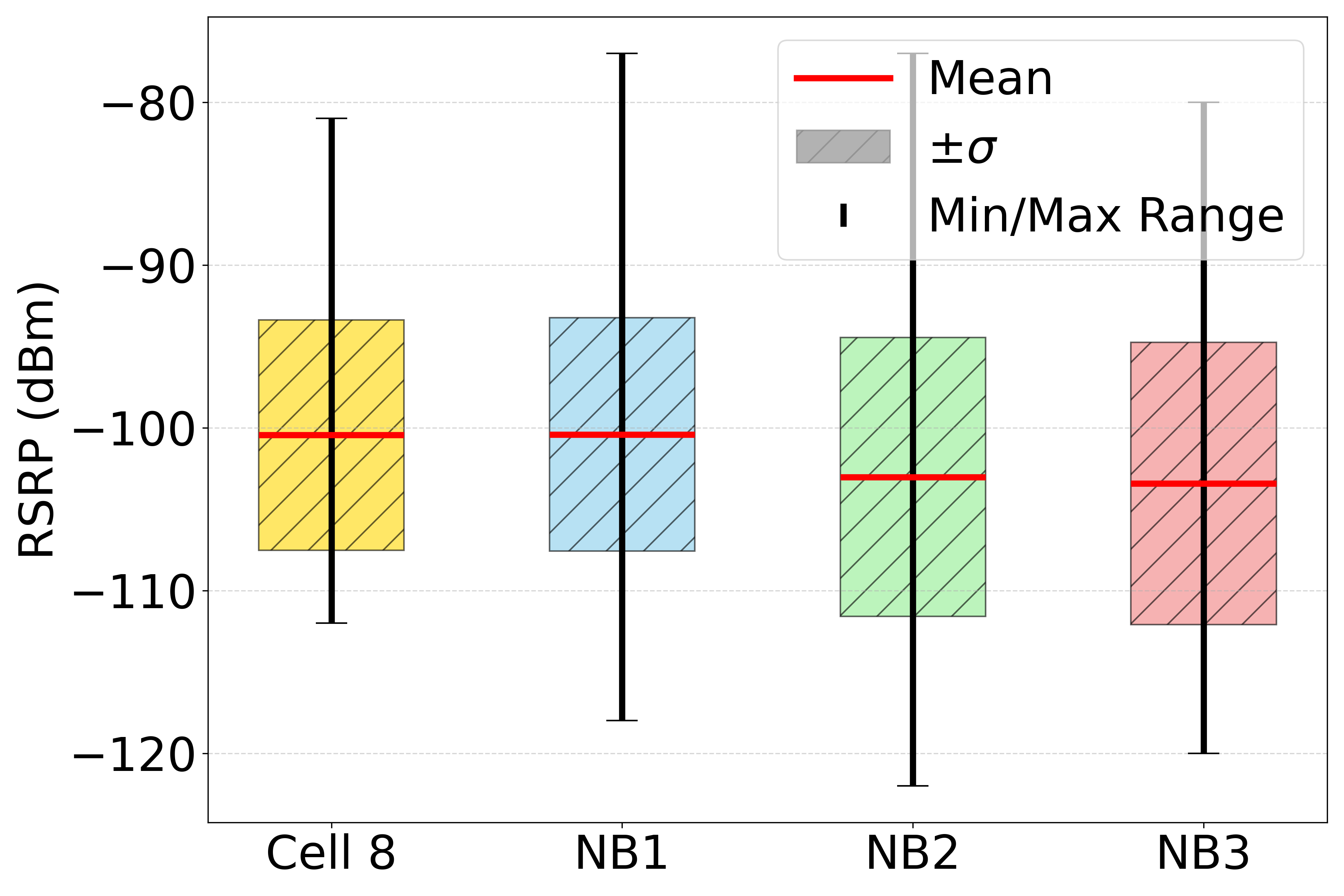}
         \vspace{-0.3cm}
        \caption{Neighbor cells RSRP.}
        \label{fig:nb_rsrp}
    \end{subfigure}\hfill
    \begin{subfigure}{0.48\linewidth}
        \centering        \includegraphics[width=.9\linewidth]{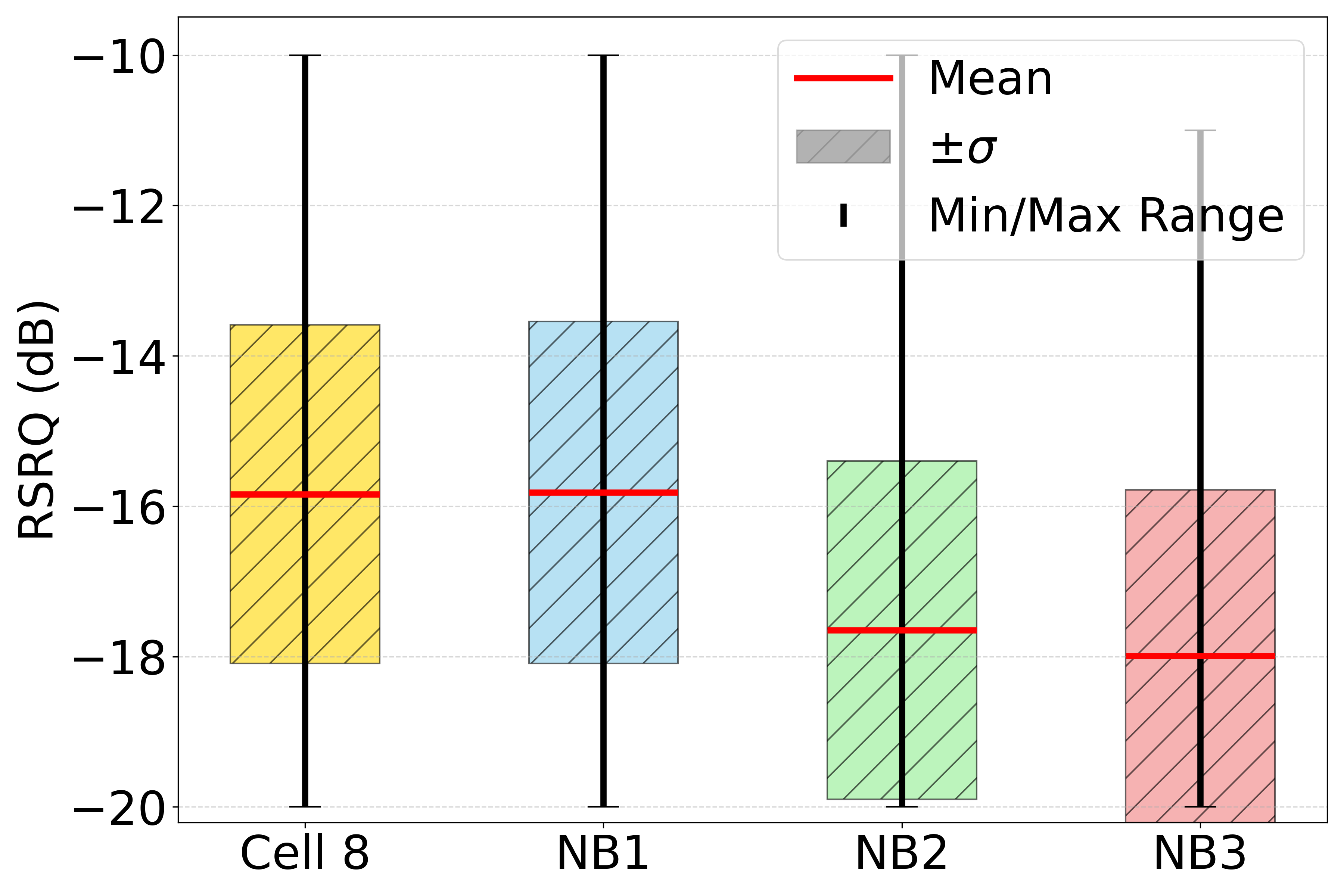}
         \vspace{-0.3cm}
        \caption{Neighbor cells RSRQ.}
        \label{fig:nb_rsrq}
    \end{subfigure}
    \vspace{-0.4cm}
    \caption{RSRP and RSRQ performance for the dominant cell (Cell 8) and three neighbor cells (denoted by ``NB1'' through ``NB3'').}
    \label{fig:nb_ran}
\end{figure}

\subsection{Neighboring Cell Performance Analysis and Handover Assessment}
\label{sec:handover}
\subsubsection{Handover readiness.} Maintaining service continuity during user mobility requires reliable handover mechanisms to neighboring cells when the serving cell signal degrades. 
The platform's multi-cell monitoring capability enables simultaneous measurement of 
neighboring cell performance, which is essential for evaluating network reliability, 
coverage redundancy, and the system's capability to maintain seamless connectivity across 
the deployment area. Our data logging platform continuously captures radio measurements 
from neighboring cells. For this analysis, we consider the three strongest neighboring 
cells, which constitute critical handover candidates during serving cell signal 
degradation events to prevent service interruption. Figure~\ref{fig:nb_ran} presents the RSRP and RSRQ performance metrics for the dominant cell (index $8$) along with three neighboring cells (denoted by ``NB'' in the figures). The experimental results confirm that good handover candidates are available, with at least one neighboring cell maintaining signal levels above threshold requirements (RSRP $\geq -100$ dBm, RSRQ $\geq -20$ dB)~\cite{3gpp36133, lai2024flight}. This coverage overlap ensures seamless handover execution and mitigates the risk of service interruption during mobility events.

\subsubsection{Handover Analysis.} 
We use Cell ID information for handover detection, which provides globally unique cell identification compared to Physical Cell IDs (PCIs) that can be reused across the network. Our algorithm processes time-series LTE measurements to simultaneously detect handovers (via Cell ID changes) and classify their trigger causes based on signal quality at the handover instant. Based on 3GPP standard ~\cite{3gpp36133}, Algorithm~\ref{algo} categorizes each detected handover into one of four event types based on threshold-based decision logic: \textbf{(i)} Handover Event 1 (E1) happens when neighbor cell RSRP exceeds serving cell by $\geq$3 dB. This event corresponds to Event A3 (Strength-based) in the context of 3GPP. \textbf{(ii)} Handover Event 2 (E2) corresponds to Interference (Quality-based) when RSRP is strong ($>$-$95$ dBm) but RSRQ is poor ($\leq$-$18$ dB). \textbf{(iii)} Handover Event 3 (E3) is triggered  when serving cell RSRP falls below -$110$ dBm. This event corresponds to Event A2 (Coverage-based) in the context of 3GPP. \textbf{(iv)} Handover Event 4 (E4) captures other handovers from network optimization or load balancing. The primary limitation is potential missed handovers during brief radio link failures, though our dataset shows $<$$0.1$\% null values ensuring high detection accuracy. This method detected and classified $217$ handover events during a $60$ minutes flight, which are distributed as follows: E1/Event A3 ($121$, $ 55.8$\%),  E2/Interference ($26$, $12.0$\%), E3/Event A2 ($66$, $30.4$\%), and E4/Other ($4$, $1.8$\%).

\begin{algorithm}[t]
\caption{Combined Handover Detection and Classification}
\label{algo}
\small
\begin{algorithmic}[1]
\Require Time-series: $\{(\text{CellID}_i, \text{RSRP}_i, \text{RSRQ}_i, \text{RSRP}_{neighbor,i}, t_i)\}_{i=1}^{n}$
\Ensure Set of classified handover events $\mathcal{H}$
\State $\mathcal{H} \leftarrow \emptyset$
\For{$i = 2$ to $n$}
    \If{$\text{CellID}_i \neq \text{CellID}_{i-1}$ and $\text{CellID}_i \neq \text{null}$}
        \State $\Delta \text{RSRP} \leftarrow \text{RSRP}_{neighbor,i-1} - \text{RSRP}_{i-1}$
        \If{$\Delta \text{RSRP} \geq 3$ dB}
            \State $cause \leftarrow \text{E1}$ \Comment{Corresponding to Event A3 in 3GPP.}
        \ElsIf{$\text{RSRP}_{i-1} > -95$ dBm and $\text{RSRQ}_{i-1} \leq -18$ dB}
            \State $cause \leftarrow \text{E2}$ \Comment{Capturing increased interferences.}
        \ElsIf{$\text{RSRP}_{i-1} \leq -110$ dBm}
            \State $cause \leftarrow \text{E3}$ \Comment{Corresponding to Event A2 in 3GPP.}
        \Else
            \State $cause \leftarrow \text{E4}$ \Comment{Handover due to other events.}
        \EndIf
        \State $\mathcal{H} \leftarrow \mathcal{H} \cup \{(t_i, \text{CellID}_{i-1}, \text{CellID}_i, cause)\}$
    \EndIf
\EndFor
\State \Return $\mathcal{H}$
\end{algorithmic}
\end{algorithm}

\begin{figure}[h]
    \centering    \includegraphics[width=0.75\linewidth]{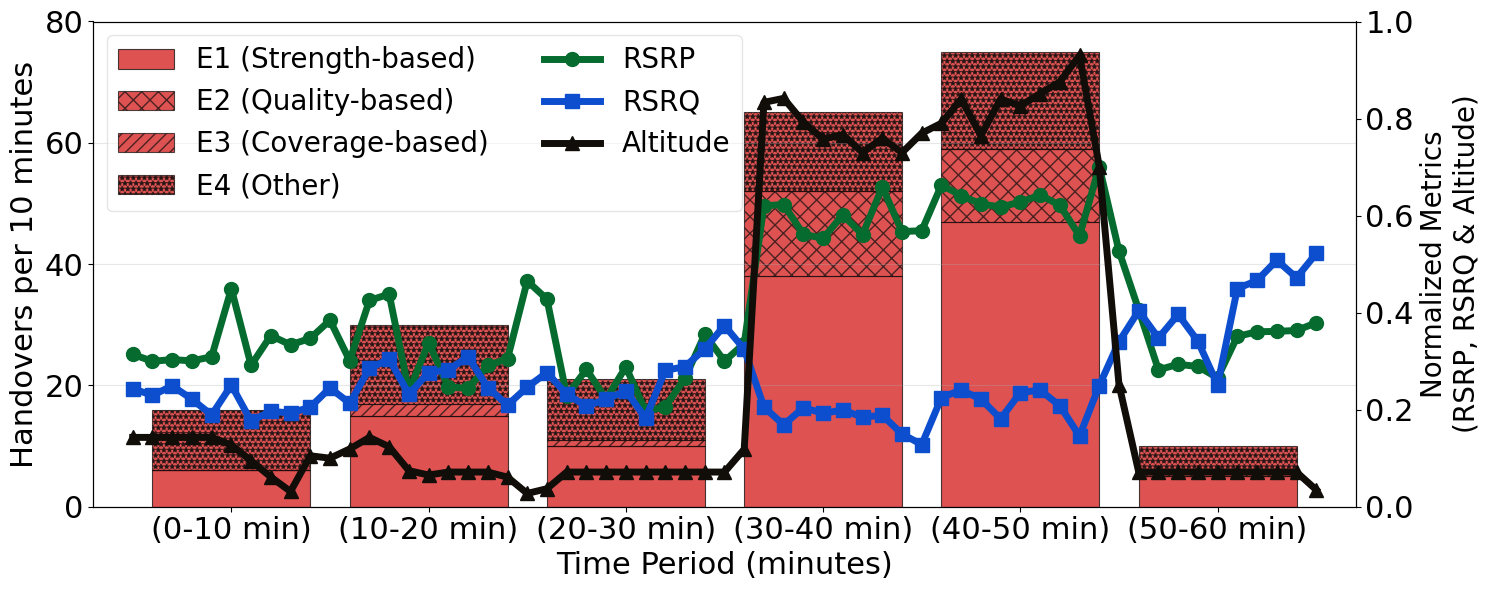}
    \vspace{-.4cm}
    \caption{Handover rate per 10-minute blocks with normalized RSRP, RSRQ, and UAV altitude.}
    \label{fig:ho_analysis}
\end{figure}

\textbf{Temporal Handover Dynamics.}
Figure~\ref{fig:ho_analysis} presents the temporal evolution of handover events over the $60$ minutes flight, with handovers aggregated into $10$-minute bins and overlaid with normalized signal quality metrics (RSRP, RSRQ) as well as flight altitude. The first $30$ minutes ($0$ - $30$ min, $250$ m ASL) show relatively low handover rates of $16$ - $30$ events per bin, with E3 (Event A2, coverage-based) handovers dominating due to weak RSRP (-$95$ to -$100$ dBm) characteristic of ground-level operations at cell edge. A significant increase in the number of handovers begins around $30$ minutes as the UAV ascends: handover rate surges to $65$ - $75$ events per $10$-minute bin during the $30$ - $50$ minute period at $330$ - $380$ m ASL, coinciding with RSRP improvement to -$81$ to -$85$ dBm, the strongest signals recorded during the entire mission. This paradox reveals that stronger signal strength does not reduce handover frequency in aerial environments. Critically, E2 (Interference) handovers emerge exclusively during this $30$ - $50$ minute high-altitude window, comprising $19$ - $22$\% of handover events as RSRQ degrades to -$16$ to -$17$ dB despite strong RSRP, indicating high co-channel interference from multiple visible cells. E1 (Event A3) handovers dominate throughout ($60$ - $70$\% of events during the $30$ - $50$ min high-altitude phase), with the $3$ dB threshold triggering frequent handover ping-pong as uniformly strong RSRP across multiple cells causes minor signal fluctuations to continuously meet the A3 criteria. The final $10$-minute bin ($50$ - $60$ min) shows handover rate returning to baseline ($10$ events) as altitude decreases to $250$ m ASL, with E3 handovers re-emerging as the primary type. This temporal analysis demonstrates that altitude, rather than signal strength, is the primary driver of handover instability, with the $30$ - $50$ minute high-altitude period exhibiting $3$ - $4$$\times$ higher handover rate than ground operations despite $15$ - $20$ dB stronger RSRP.

\begin{wrapfigure}{r}{2.2in}
    \centering   
    \vspace{-.2cm}
    \includegraphics[width=\linewidth]{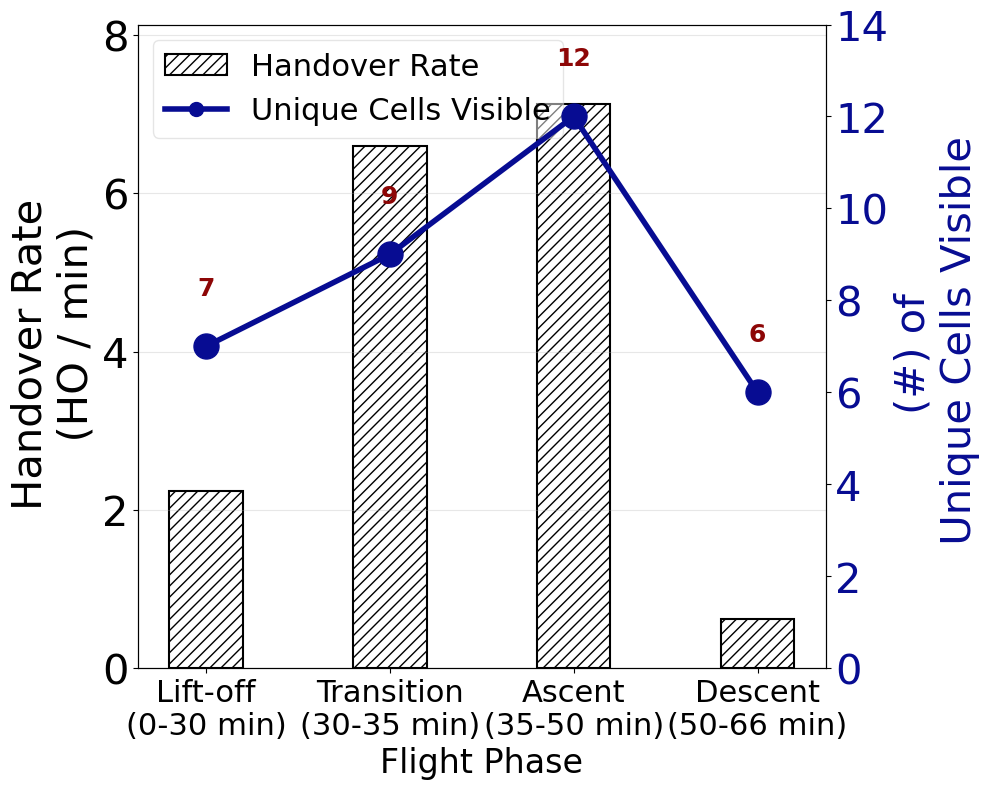}
    \vspace{-.9cm}
    \caption{Handover rate change and cell visibility across flight phases.}
    \label{fig:ho_cells}
    \centering
\end{wrapfigure}

\textbf{Cell Visibility and Network Reporting.}
Figure~\ref{fig:ho_cells} examines the relationship between unique cell visibility and handover rate across four flight phases: Lift-off ($0$ - $30$ min, ground level at $250$ m ASL), Transition ($30$ - $35$ min, rapid ascent from $250$ - $330$ m ASL), Ascent ($35$ - $50$ min, sustained flight at $330$ - $380$ m ASL), and Descent ($50$ - $66$ min, return to $250$ m ASL). During the Lift-off phase, the UAV detects $7$ unique Cell IDs with a handover rate of $2.2$ HO/min, limited by terrain obstructions and typical ground-level propagation. The Transition phase (converting to fixed-wing), marking the beginning of vertical ascent, experiences a $3$$\times$ surge in handover rate to $6.6$ HO/min with a slight increase in cell visibility ($9$ Cell IDs), demonstrating that rapid altitude change velocity alone drives handover instability through dramatic RSRP fluctuations. The Ascent phase reveals the most striking finding: cell visibility increases $33$\% to $12$ unique Cell IDs due to LOS propagation, enabling detection of cells $5$ - $10$ km away, yet handover rate increases disproportionately by $223$\% to $7.1$ HO/min, compared to ground level. The increased number of visible cells could have led to suboptimal and frequent handover decisions, which may not necessarily be due to performance degradation received from the serving cell~\cite{3gpp36331, tashan2022mobility, lobinger2010load}.
The Descent phase shows handover rate dropping to $0.6$ HO/min with $6$ visible Cell IDs as the UAV returns to ground level, confirming that both altitude and vertical velocity are key factors in handover behavior.

\begin{wrapfigure}{r}{2.8in}
    \centering
    \vspace{-.32cm}
    \includegraphics[width=\linewidth]{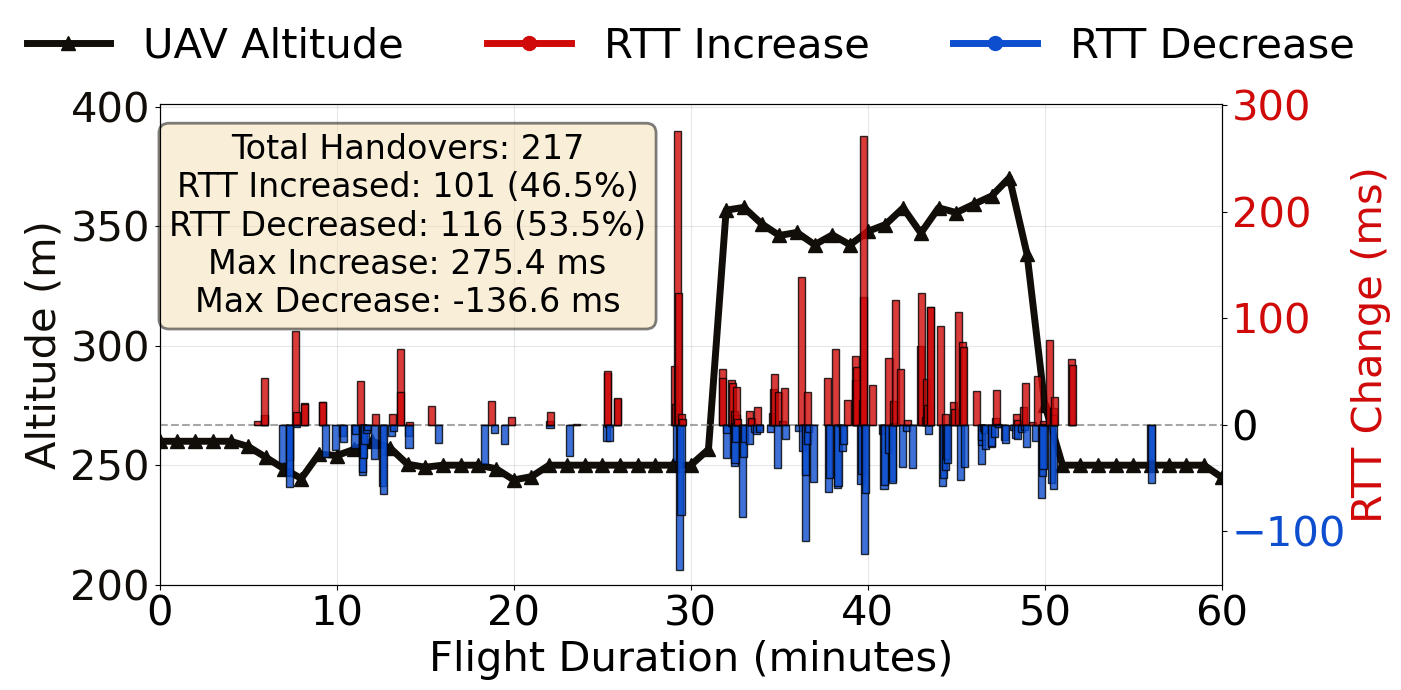}
    \vspace{-.65cm}
    \caption{Changes in RTT at Handover (HO) events overlaid on UAV altitude trajectory.}
    \label{fig:ho_rtt}
\end{wrapfigure}
\textbf{RTT Impact of Handover Events.}
Figure~\ref{fig:ho_rtt} overlays individual handover RTT impacts on the UAV altitude trajectory, revealing that handovers do not uniformly degrade latency as commonly assumed. Analysis of all $217$ handovers shows that $116$ events ($53.5$\%) decreased RTT while $101$ events ($46.5$\%) increased RTT, with a maximum RTT increase of $275.4$ ms and maximum RTT decrease of -$136.6$ ms. During the first $30$ minutes at ground level ($250$ m ASL), RTT changes are moderate and mixed, with roughly equal numbers of improvements and degradations typically ranging $\pm$$50$ ms. 
The most significant RTT instability occurs during the $30$ - $50$ minute high-altitude phase ($330$ - $380$ m ASL), where both the largest RTT increases ($100$ - $275$ ms spikes) and substantial decreases ($50$ - $100$ ms improvements) are concentrated. 
This period, which coincides with peak handover frequency ($7.1$ HO/min), strongest RSRP (-$81$ to -$85$ dBm), and degraded RSRQ (-$16$ to -$17$ dB), demonstrates that strong signal strength alone cannot prevent latency degradation. The RTT impact exhibits critical asymmetry: while $53.5$\% of handovers improve latency (often by moving the UE from congested or poorly-backhauled cells to better network conditions), worst-case degradation ($275.4$ ms) is $2.0$~$\times$ larger than best-case improvement ($136.6$ ms), creating a high-risk profile where handovers either help modestly (typical decrease: $20$ - $50$ ms) or hurt severely (outliers exceeding $100$ - $200$ ms). The descent phase ($50$ - $60 $ min, returning to $250$ m ASL) shows sparse RTT events reflecting the low handover rate ($0.6$ HO/min), with mixed but moderate impacts and no extreme outliers. At the sustained $7.1$ HO/min rate observed during the high-altitude phase (one handover every $8.5$ seconds), applications experience continuous latency disruptions with insufficient time for recovery between events, resulting in persistent performance degradation unsuitable for time-sensitive UAV operations such as real-time video streaming and command-and-control.

\textbf{Key Findings and Implications.} Our handover analysis establishes the following key findings: 
\begin{itemize} [leftmargin = 4mm, itemsep = 0.009in, parsep = 0.005in, topsep = 0.005in]
    \item \textbf{Altitude drives 3-4$\times$ handover rate increase despite $15$ - $20$ dB stronger RSRP,} revealing that RSRP-only algorithms fail in aerial LOS conditions where uniformly strong signals from multiple cells trigger excessive E1 (Event A3) handovers.
    
    \item \textbf{Interference handovers appear exclusively during the high-altitude phase ($30$ - $50$ min at $330$ - $380$ m ASL)} where strong RSRP coexists with poor RSRQ (-$16$ to -$17$ dB), with E2 (Interference) events comprising $12$\% of total handovers. This demonstrates that quality metrics must supplement strength-based handover triggers for aerial platforms.
    
    
    \item \textbf{Handover impact is asymmetric:} $53.5$\% of handovers improve RTT, but worst-case degradation ($275$ ms) is $2$ $\times$ larger than best-case improvement ($137$ ms), creating high-risk latency profiles unsuitable for time-sensitive applications.
\end{itemize}

\section{Cellular vs. Starlink: End-to-End Service Quality Assessment}
\label{sec:comparison}
Given the platform's ability to simultaneously track metrics from both 
terrestrial cellular network and LEO satellite (Starlink Mini) connectivity, 
we perform dual connectivity analysis comparing end-to-end performance 
characteristics under operational flight conditions. The analysis focuses on 
three critical metrics: round-trip latency, bidirectional throughput, and 
packet delivery statistics.
\subsection{Head-to-Head Comparisons}

\begin{figure}[!t]
    \centering
    \begin{subfigure}{0.32\linewidth}
        \centering
        \includegraphics[width=\linewidth]{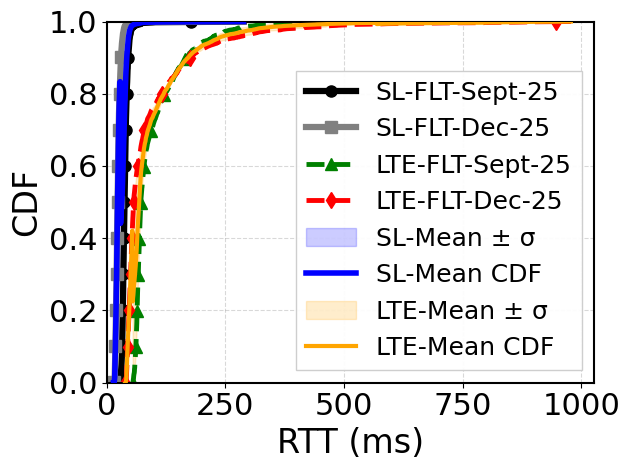}
        \vspace{-0.6cm}
        \caption{RTT.}
        \label{fig:pmlte_star_rtt}
    \end{subfigure}
    \begin{subfigure}{0.32\linewidth}
        \centering        \includegraphics[width=\linewidth]{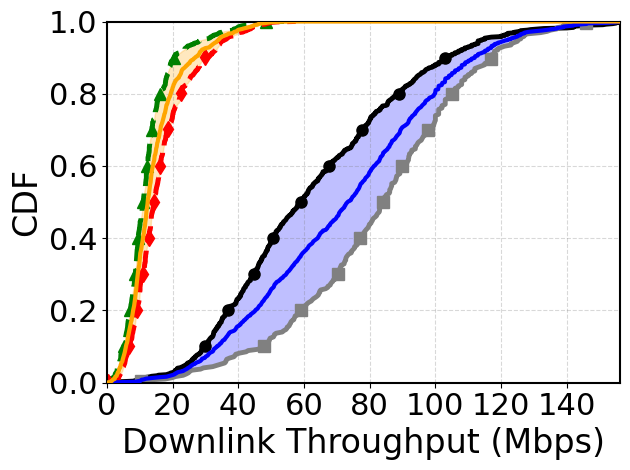}
        \vspace{-0.6cm}
        \caption{Downlink Throughput.}
        \label{fig:pmlte_star_dlthrpt}
    \end{subfigure}
    \begin{subfigure}{0.32\linewidth}
        \centering
        \includegraphics[width=\linewidth]{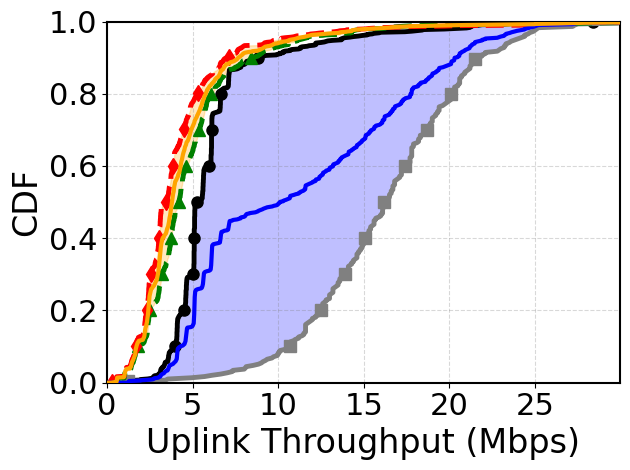}
        \vspace{-0.6cm}
        \caption{Uplink Throughput.}
        \label{fig:pmlte_star_ulthrpt}
    \end{subfigure}
    \vspace{-0.4cm}
    \caption{End-to-end network performance measurement. The LTE and Starlink (denoted by ``SL'')  performance were measured simultaneously at identical times and locations (between UAV and a dedicated remote server). The experiments were repeated two times in September and December 2025.}
    \label{fig:e2eLTEStar}
\end{figure}

\textbf{Latency Performance:} We perform comparative analysis of LTE and Starlink performance measured simultaneously at identical times and locations (between UAV and a dedicated remote server). The comparison was repeated over two flight tests (September 2025 and December 2025), and the results are shown in Figure~\ref{fig:e2eLTEStar}. The operational flight tests revealed significant 
performance differences in real-time responsiveness between the two connectivity 
options. As shown in Figure~\ref{fig:pmlte_star_rtt}, the Starlink satellite 
link demonstrated exceptional latency stability, with $95\%$ of RTT measurements 
remaining below $50$ ms. In contrast, the cellular network exhibited 
higher latency, with approximately $80\%$ of measurement data falling 
under the $150$ ms threshold. This three-fold difference in latency distribution 
indicates that the LEO satellite constellation provides superior responsiveness 
for real-time applications despite the additional propagation distance to orbit.

\textbf{Throughput Performance:} The throughput analysis reveals 
performance dichotomies between cellular and Starlink across downlink and uplink 
directions. For downlink performance, as shown in Figure~\ref{fig:pmlte_star_dlthrpt}, 
Starlink demonstrates a significant capacity advantage with $95\%$ of measurements 
exceeding $25$ Mbps. The cellular network shows more modest performance, with the 
comparable $95\%$ threshold at only $5$ Mbps. This five-fold difference in 
downlink capacity highlights the bandwidth advantages of the LEO satellite system 
for data-intensive applications. However, uplink performance presents a contrasting 
picture. Approximately $65\%$ of uplink measurements for both cellular and Starlink 
fall below $5$ Mbps, indicating similar uplink constraints across both technologies. 
This symmetric uplink limitation suggests that uplink capacity represents a shared 
bottleneck regardless of the underlying connectivity technology.

\textbf{Packet Delivery Statistics:} Both connectivity solutions demonstrated 
high transmission reliability during active flight operations. As shown in 
Table~\ref{tab:packet_delivery}, the cellular connection achieved 
a packet delivery rate of $99.44\%$ ($5130$ out of $5159$ packets delivered), while the 
Starlink connection maintained a $99.31\%$ delivery rate ($3602$ of $3627$ packets 
delivered). These minimal packet loss rates ($0.56\%$ for cellular and $0.69\%$ for 
Starlink) validate the robustness of both links for UAV operations and indicate 
sufficient reliability for data transfers. 


\begin{table}[ht]
    \centering

    \begin{tabular}{lcccc}
        \toprule
        \textbf{Connection} & \textbf{Pkts Sent} & \textbf{Pkts Delivered} & \textbf{Pkts Dropped} & \textbf{Delivery Rate (\%)} \\
        \midrule
        LTE      & 5159 & 5130 & 29 & 99.44 \\
        Starlink & 3627 & 3602 & 25 & 99.31 \\
        \bottomrule
    \end{tabular}
        \caption{Packet delivery performance comparison between LTE and Starlink connections.}
    \label{tab:packet_delivery}
\end{table}

\begin{figure}[!t]
\centering
\begin{subfigure}{0.32\linewidth}
    \centering
  \includegraphics[width=\linewidth]{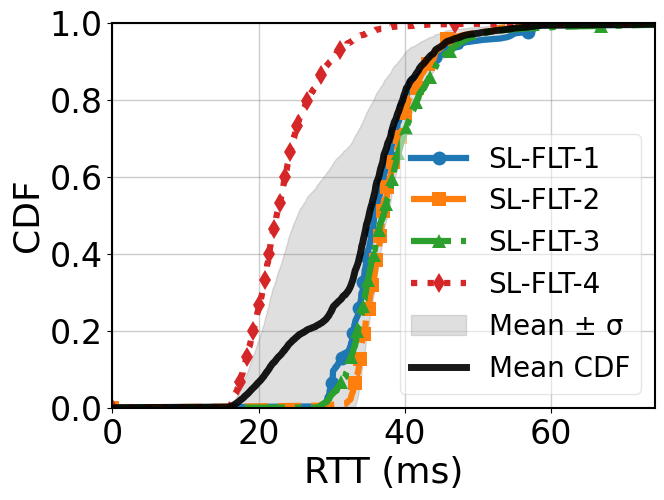}
  \vspace{-0.65cm}
  \caption{Starlink RTT.}
  \label{fig:star_temp_rtt}
  \end{subfigure}
 \begin{subfigure}{0.32\linewidth}
    \centering
 \includegraphics[width=\linewidth]{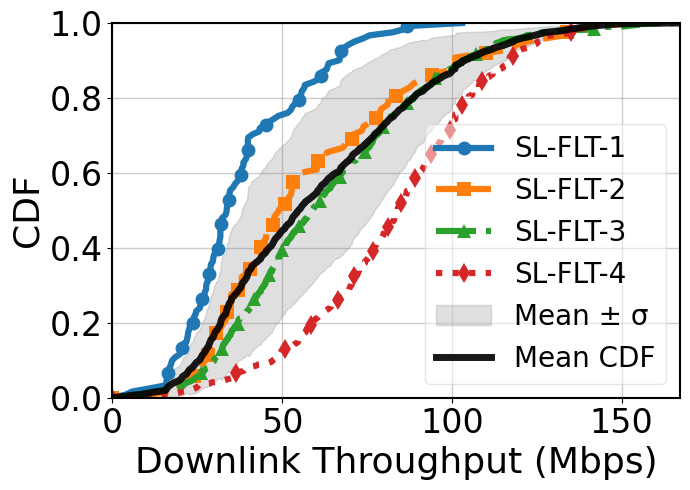}  
  \vspace{-0.6cm}
  \caption{Starlink Downlink Throughput.}
  \label{fig:star_temp_dlthrpt}
  \end{subfigure}
  \begin{subfigure}{0.32\linewidth}
 \centering
 \includegraphics[width=\linewidth]{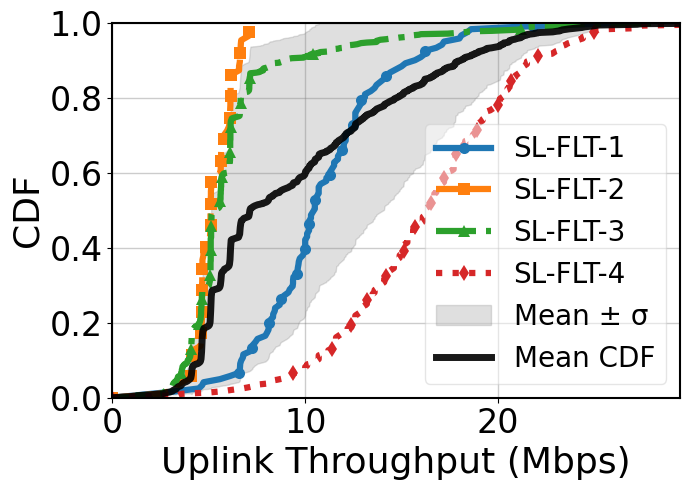}  
  \vspace{-0.6cm}
  \caption{Starlink Uplink Throughput.} 
  \label{fig:star_temp_ulthrpt}
  \end{subfigure}
   \vspace{-0.4cm}
  \caption{Temporal stability analysis of end-to-end metrics for Starlink across four distinct flight tests (denoted by ``FLT-1'' through ``FLT-4'') conducted over a period of three months.}
\label{fig:e2e_temp_Star}
\end{figure}

\textbf{Dual Connectivity Implications:} The complementary performance characteristics of cellular and Starlink suggest opportunities for intelligent dual connectivity architectures. Starlink's latency performance ($95\%$ under $50$ ms) makes it well-suited for latency-sensitive control and telemetry traffic, while its high downlink capacity ($95\%$ exceeding $25$ Mbps) supports bandwidth-intensive payload data transmission. The cellular network, while exhibiting higher latency, maintains competitive packet delivery reliability ($99.66\%$) and provides 
consistent uplink performance comparable to Starlink. A dual connectivity approach could leverage Starlink for primary data transmission and real-time operations while utilizing cellular for redundancy and failover, or alternatively, could intelligently distribute traffic based on latency requirements and bandwidth availability. The platform's capability to simultaneously measure both links during flight operations enables real-time link quality assessment necessary for dynamic connectivity management in dual-link architectures.

\subsection{Starlink Performance Deep Dive}
\label{sec:slflight}
\subsubsection{Short-term Starlink Performance Evolution.} 
\begin{figure}[t]
    \centering
    \begin{subfigure}{0.49\linewidth}
        \centering
        \includegraphics[width=\linewidth]{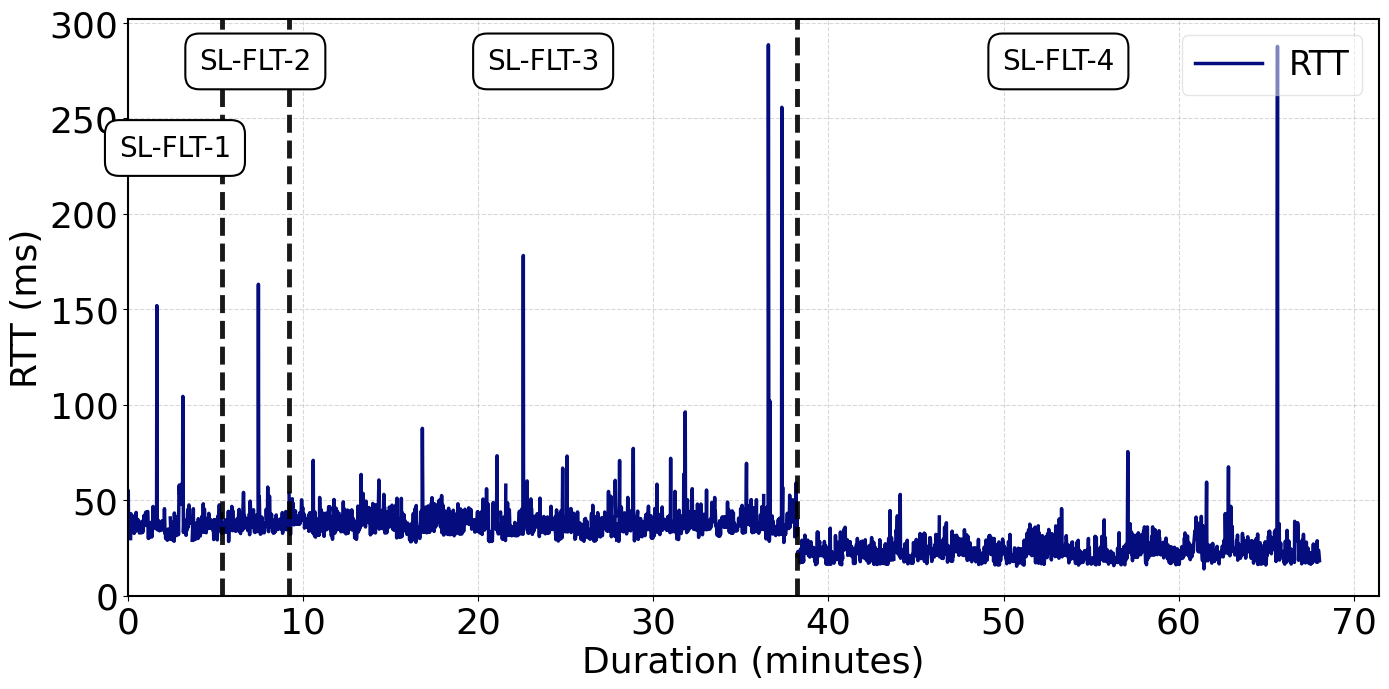}
         \vspace{-0.6cm}
        \caption{RTT.}
        \label{fig:st_rtt_temp_total}
    \end{subfigure}
    \begin{subfigure}{0.49\linewidth}
        \centering
        \includegraphics[width=\linewidth]{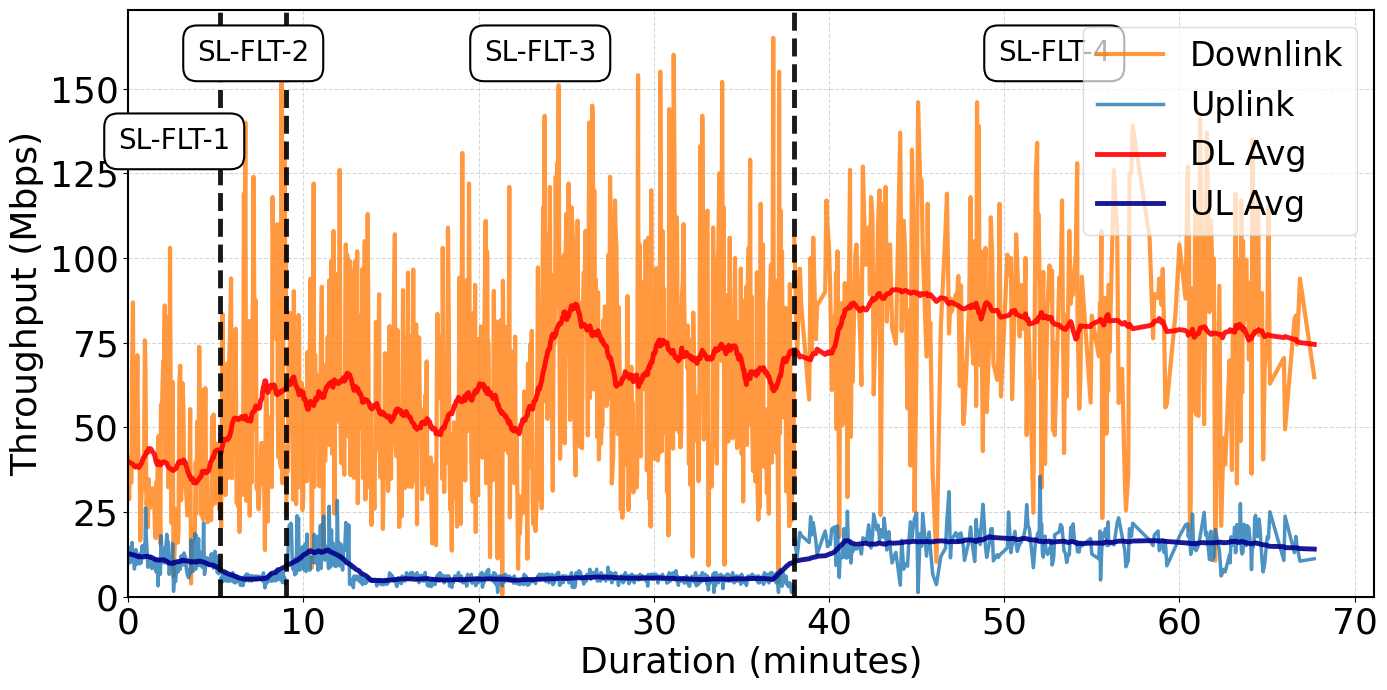}
         \vspace{-0.6cm}
        \caption{Uplink and downlink throughput.}
        \label{fig:st_bidr_temp_total}
    \end{subfigure}
     \vspace{-0.4cm}
    \caption{Starlink E2E metrics conducted across 4 flight tests to identify any spikes similar to ground sensitivity analysis as shown in Figure~\ref{fig:ground_test_results}. }
    \label{fig:sky_test_results}
\end{figure}

In this part, we conduct more comprehensive performance analysis on Starlink network performance using our UAV measurement platform. We repeat the  measurements across four different flight tests (denoted by ``SL-FLT-1'' through ``SL-FLT-4'') over a $3$-month period (from September 2025 to December 2025). Figures \ref{fig:e2e_temp_Star} and \ref{fig:sky_test_results} demonstrate the CDF and temporal performance across the four flight tests. From the results, we observe several interesting trends. As shown in Figure~\ref{fig:sky_test_results}, the concatenated time series plots reveal a clear temporal performance evolution in Starlink connectivity from September to December 2025. SL-FLT-4 conducted on Dec 17, 2025, demonstrates substantially superior performance across all metrics compared to earlier test days. Specifically, December shows $\sim$ $5$$\times$ higher downlink throughput (consistently $60$ - $100$ Mbps vs $20$ - $40$ Mbps in September), $\sim$ $3$$\times$ higher uplink throughput ($10$ - $20$ Mbps vs $5$ - $8$ Mbps), and ~$40$\% lower RTT ($20$ - $30$ ms vs $40$ - $60$ ms). Additionally, December data exhibits dramatically reduced variability as the performance curves are smoother and more stable, contrasting sharply with the erratic fluctuations visible in SL-FLT-1 and SL-FLT-2. This aligns perfectly with the CDF analysis (Figure~\ref{fig:e2e_temp_Star}), where we observed that the December distribution was better with mean improvements of +$30$\% downlink, +$160$\% uplink, and -$39$\% latency. Furthermore, in contrast to the ground-based sensitivity analysis depicted in Figure~\ref{fig:ground_test_results}, our observations revealed no unusual disruptions, with the exception of a handful of transient spikes. This configuration of Starlink antennas on the UAV likely enhanced LOS conditions, ensuring that connectivity remained stable even amid the UAV's maneuvers. Based on the observed performance, we concur that Starlink's LEO connectivity provides a reliable communication channel for real-time applications involving UAVs in rural settings.

\subsubsection{Impacts of UAV Dynamics on Starlink performance}
During flight test exercise, the UAV followed a race-track trajectory involving continuous banking, pitching, and turning. We observed that despite the frequent changes in the Starlink antenna's physical orientation caused by aircraft banking and pitching, no significant impact on link stability was observed. Our statistical analysis confirms relatively small correlations between the aircraft’s altitude angles and network performance. The correlation coefficients between RTT and roll/pitch were calculated to be $0.0434$ and $-0.1122$, respectively. Similarly, the correlation coefficients between throughput and roll/pitch were $-0.0011$ and $0.0544$. These actual flight test results indicate that the Starlink Mini’s phased-array beam-steering is sufficiently robust for standard fixed-wing UAV operations assuming the antenna maintained a general skyward orientation.

\section{ML-based 3D Performance Prediction}
\label{sec:ml}
As discussed in earlier sections, the platform captures measurements at various altitudes during flight tests, providing comprehensive spatial sampling across the deployment area. 
However, exhaustive measurement of every possible location and altitude combination is impractical. We leverage \emph{standard} ML models to predict performance metrics at unmeasured altitudes and locations. Our approach involves training these models on GPS-tagged position 
data from flight tests, where each measurement is associated with a specific location defined by longitude, latitude, and altitude. We use position coordinates as input features 
($x$) to predict corresponding metric values as outputs ($y$). Once trained, the models accept new position coordinates ($\hat{x}$) to generate predicted values ($\hat{y}$).

We use two distinct evaluation paradigms: Leave-One-Altitude-Out (LOAO) cross-validation and the ``standard'' $80$/$20$ Random Split approach. The LOAO strategy is specifically designed to test the models' vertical extrapolation capabilities by removing all measurements from a target altitude during training and evaluating the model on that unseen vertical plane. 
Specifically, we create a truncated dataset by removing all measurements from one altitude level from the training set and train each model on the remaining altitude measurements. The trained model was then used to predict metric values for the excluded altitude. 
In contrast, the 
$80$/$20$ Random Split serves as a benchmark for general spatial interpolation, where $80\%$ of the shuffled dataset is used for training and the remaining $20\%$ for testing across all spatial coordinates and altitudes simultaneously. 
This evaluation framework enables a comprehensive assessment of both altitude-specific extrapolation and general spatial prediction capabilities.

\textbf{Evaluation Metrics.}
To evaluate model performance, we used Mean Absolute Error (MAE) as the primary evaluation metric and Root Mean Square Error (RMSE) as a secondary metric. These metrics are computed as: 
\begin{equation}\label{eq:MAE}
    \text{MAE} = \frac{1}{N} \sum_{i=1}^{N} |y_i - \hat{y}_i|, \ \ \  \text{and} \ \ \ \text{RMSE} = \sqrt{\frac{1}{N} \sum_{i=1}^{N} (y_i - \hat{y}_i)^2}, 
\end{equation}
where $N$ is the number of observations, and $y_i$ and  $\hat{y}_i$ are the actual (collected) and predicted values for the $i^{th}$ observation, respectively. Unlike MAE, which treats all errors equally, RMSE assigns greater importance to larger errors due to the squaring operation, making it particularly useful for identifying models that produce occasional large prediction errors.
    

\textbf{Results and Discussion.} Experimental results in Table~\ref{tab:ml_eval} demonstrate that while both training approaches achieve high fidelity, the models exhibit significantly higher precision when performing spatial interpolation compared to vertical extrapolation. As shown in Table~\ref{tab:ml_eval}, the Random Forest (RF) model achieved its peak accuracy under the $80$/$20$ split with an RMSE of $1.764$ dBm and an MAE of $1.258$ dBm. However, when subjected to the more challenging LOAO task, the RMSE increased to $3.610$ dBm. This ``extrapolation penalty'' quantifies the additional variance introduced when predicting signal metrics at entirely unobserved altitudes, which is a critical consideration for network operators managing $3$D aerial coverage in rural terrain.
\begin{table}[!t]
\centering
\begin{tabular}{clcccc}
\hline
\multirow{2}{*}{\textbf{RAN Metric}} & \multirow{2}{*}{\textbf{Machine Learning Model}} & \multicolumn{2}{c}{\textbf{RMSE (dBm)}} & \multicolumn{2}{c}{\textbf{MAE (dBm)}} \\
\cline{3-4} \cline{5-6}
& & \textit{LOAO} & \textit{Generic 80-20} & \textit{LOAO} & \textit{Generic 80-20} \\
\hline
\multirow{3}{*}{RSRP} & Random Forest (RF) & 5.21 & 3.97 & 3.96 & 2.43 \\
& Gradient Boosting (GB) & 5.14 & 4.15 & 3.92 & 2.78 \\
& Multi-Layer Perceptron (MLP) & 5.88 & 4.61 & 4.73 & 3.36 \\
\hline
\multirow{3}{*}{RSRQ} & Random Forest (RF) & 2.24 & 1.84 & 1.85 & 1.45 \\
& Gradient Boosting (GB) & 2.27 & 1.90 & 1.86 & 1.52 \\
& Multi-Layer Perceptron (MLP) & 2.57 & 2.12 & 2.11 & 1.76 \\
\hline
\multirow{3}{*}{RSSI} & Random Forest (RF) & 4.58 & 3.32 & 3.50 & 2.40 \\
& Gradient Boosting (GB) & 4.47 & 3.50 & 3.46 & 2.59 \\
& Multi-Layer Perceptron (MLP) & 4.84 & 3.79 & 3.77 & 2.89 \\
\hline
\multirow{3}{*}{SINR} & Random Forest (RF) & 2.44 & 1.68 & 1.85 & 1.12 \\
& Gradient Boosting (GB) & 2.54 & 1.80 & 1.92 & 1.24 \\
& Multi-Layer Perceptron (MLP) & 2.65 & 2.09 & 2.11 & 1.56 \\
\hline
\end{tabular}
\caption{Performance Comparison of Machine Learning Models for predicting the RAN metrics with 2 methods Leave-One-Altitude-Out (LOAO) and the generic $80/20\%$ train/test split method. }
\label{tab:ml_eval}
\end{table}
The performance gap between the two methodologies highlights the complexity of the characterization of the $3$D radio signal. 
For safety-critical UAV operations, the LOAO metrics (specifically the RMSE of about $3.6$ dBm) provide the worst-case error margin for autonomous path planning at new flight levels. These findings confirm that while ML-based interpolation is highly precise, vertical extrapolation requires slightly higher safety buffers to account for altitude-dependent propagation phenomena like diffraction and ground reflections.

These predictive capabilities have direct implications for the optimization of BVLOS UAV operations and cellular network planning. By enabling accurate 3D coverage extrapolation, the models allow network operators to estimate interference patterns and signal dead zones at altitudes where physical measurements are logistically infeasible. Furthermore, the integration of these models into autonomous flight control systems supports proactive path optimization. A UAV can leverage these signal predictions to dynamically adjust its trajectory, maintaining a reliable communication link by avoiding predicted zones of high multi-cell interference or weak signal power, thereby enhancing mission safety and data transmission reliability.

\section{Conclusion}
\label{sec:conclusion}
In this paper, we presented the first open-source, dual-connectivity UAV-based measurement platform to characterize the performance of terrestrial LTE and LEO satellite networks in rural environments.  
Our measurement platform integrates commercial-grade cellular modem and Starlink Mini terminal with  onboard flight controller and computing systems to capture synchronized physical, network, and application-layer metrics across diverse flight altitudes. 
Experimental validation demonstrates the platform's capability to reveal altitude-dependent performance trade-offs: signal power improves $8-20$ dB at higher altitudes due to enhanced LOS conditions, whereas signal quality degrades $3-5$ dB because of increased interference from neighboring cells. This causes $3-4$ $\times$ increase in handover
rates due to excessive multi-cell visibility rather than signal degradation. 
Contrary to the general belief that satellite links are slower than terrestrial networks,
our LTE-Starlink comparative analysis showed that Starlink maintained $95$\% of RTT samples below $50$ ms, significantly outperforming the rural LTE network with $20\%$ higher than $150$ ms RTT.  Finally, our ML-based prediction suite quantifies the ``extrapolation penalty'' inherent in 3D network measurements. 
To facilitate reproducible research, we will make the complete platform design, software suite, and spatiotemporal dataset publicly available. As the next steps, we plan to extend the platform capabilities to support 5G NR measurements and multi-UAV measurements across larger geographic areas. Furthermore, we will extend our measurement campaigns to other geographical environments and with multiple cellular network operators. 
\newpage
\bibliographystyle{IEEEtran}
\bibliography{ref}

\end{document}